\documentclass[%
superscriptaddress,
preprint,
nofootinbib,
nobibnotes,
 amsmath,amssymb,
 aps,
 prd
]{revtex4}

\usepackage{graphicx}
\usepackage{dcolumn}
\usepackage{bm}

\newcommand{\qed}{\nobreak \ifvmode \relax \else
      \ifdim\lastskip<1.5em \hskip-\lastskip
      \hskip1.5em plus0em minus0.5em \fi \nobreak
      \vrule height0.75em width0.5em depth0.25em\fi}
      
\begin{document}

\preprint{}

\title{A Semi-Parametric Approach to the Detection of Non-Gaussian
Gravitational Wave Stochastic Backgrounds}

\author{Lionel Martellini}
 \email{lionel.martellini@edhec-risk.com}
\affiliation{EDHEC-Risk Institute, 400 Promenade des Anglais, BP 3116, 06202 Nice
Cedex 3, France} 
\affiliation{UMR ARTEMIS, CNRS,
University of Nice Sophia-Antipolis, Observatoire de la C\^{o}te d'Azur, CS 34229 F-06304 NICE, France}
\author{Tania Regimbau}
\affiliation{UMR ARTEMIS, CNRS,
University of Nice Sophia-Antipolis, Observatoire de la C\^{o}te d'Azur, CS 34229 F-06304 NICE, France}

\date{\today}

\begin{abstract}
Using a semi-parametric approach based on the fourth-order Edgeworth expansion for the unknown signal distribution, 
we derive an explicit expression for the likelihood detection statistic in the presence of non-normally distributed gravitational wave
stochastic backgrounds. Numerical likelihood maximization exercises based on Monte-Carlo
simulations for a set of large tail symmetric non-Gaussian distributions suggest that the fourth cumulant of the signal distribution can be estimated with reasonable precision when the ratio between the signal and the noise variances is larger than 0.01. The estimation of higher-order cumulants of the observed gravitational wave signal distribution is expected to provide additional constraints on astrophysical and cosmological models.

\end{abstract}

\pacs{Valid PACS appear here}
\maketitle

\section{\label{sec:intro} Introduction}

According to various cosmological scenarios, we are bathed in a stochastic background of gravitational waves. Proposed theoretical models include the amplification of vacuum fluctuations during inflation\cite{1975JETP...40..409G,1993PhRvD..48.3513G,1979JETPL..30..682S}, pre Big Bang models \cite{1993APh.....1..317G,1997PhRvD..55.3330B,2010PhRvD..82h3518D}, cosmic (super)strings \cite{2005PhRvD..71f3510D,2007PhRvL..98k1101S,2010PhRvD..81j4028O,2012PhRvD..85f6001R} or phase transitions \cite{2008PhRvD..77l4015C,2009PhRvD..79h3519C,2009JCAP...12..024C}. In addition to the cosmological background (CGB) \cite{2000PhR...331..283M,2012JCAP...06..027B}, an astrophysical contribution (AGB) \cite{2011RAA....11..369R} is expected to result from the superposition of a large number of unresolved sources, such as core collapses to neutron stars or black holes \cite{2005PhRvD..72h4001B,2006PhRvD..73j4024S,2009MNRAS.398..293M,2010MNRAS.409L.132Z}, rotating neutron stars \cite{2001A&A...376..381R,2012PhRvD..86j4007R} including magnetars \cite{2006A&A...447....1R,2011MNRAS.410.2123H,2011MNRAS.411.2549M,2013PhRvD..87d2002W}, phase transition \cite{2009GReGr..41.1389D} or initial instabilities in young neutron stars \cite{1999MNRAS.303..258F,2011ApJ...729...59Z,2004MNRAS.351.1237H,2011ApJ...729...59Z} or compact binary mergers \cite{2011ApJ...739...86Z,2011PhRvD..84h4004R,2011PhRvD..84l4037M,2012PhRvD..85j4024W,2013MNRAS.431..882Z}. 
Many models are within the reach of the next generation of GW detectors such as Ad. LIGO/Virgo \cite{AdLIGO,AdVIRGO}. In particular compact binary coalescences have a good chance to be detected after 3-4 years of operation \cite{2012PhRvD..85j4024W}. With the third generation detector Einstein Telscope \cite{ET}, these models could be detected with a large signal-to-noise ratio allowing for parameter estimation. The detection of the cosmological background would give very important constraints on the first instant of the Universe, up to the limits of the Planck era and the Big Bang, while the detection of the AGB would provide crucial information about the star formation history, the mass range of neutron star or black hole progenitors or the rate of compact binary mergers. 

The optimal detection strategy to search for a stochastic background is to cross correlate the output of two detectors (or of a network of detectors) to eliminate the instrumental noise \cite{1999PhRvD..59j2001A}. The GW background is usually assumed to be Gaussian invoking the central limit theorem , and thus completely characterized by its mean and variance. However recent predictions based on population modeling suggest that for many astrophysical models, there may not be enough overlapping sources, resulting in the formation of a non-Gaussian background. It has also been shown that the background from cosmic strings could be dominated by a non-Gaussian contribution arising from the closest sources \cite{2005PhRvD..71f3510D,2012PhRvD..85f6001R} . 
The identification of a non-Gaussian signature would not only permit to distinguish between astrophysical and cosmological GW backgrounds and gain confidence in a detection, but the measurement of extra information would also improve parameter estimation and our understanding of the models.

In the past decade a few methods have been proposed to search for a non-Gaussian stochastic background, including the \textit{probability horizon} concept developed by \cite{2005MNRAS.361..362C} based on the temporal evolution of the loudest detected event on a single detector, the maximum likelihood statistic of \cite{2003PhRvD..67h2003D} which extends the standard cross correlation statistic in the time domain in the case of short astrophysical bursts separated by long periods of silence, the fourth-order correlation method \cite{2009PhRvD..80d3003S} which uses fourth-order correlation between four detectors to measure the third and the fourth moments of the distribution of the GW signal, or the recent extension of the standard cross-correlation statistic by \cite{2013PhRvD..87d3009T}.

In this paper, we start from the general formalism presented in \cite{2003PhRvD..67h2003D} (DF03) to analyze small deviations from the Gaussian distribution. In this case the cross-correlation statistic is almost optimal and allows for the estimation of the variance of the signal distribution, but it cannot be used to estimate higher order moments.
The approach we propose is based on Edgeworth expansion, which is a formal asymptotic expansion of the characteristic function of the signal distribution, whose unknown
probability density function is to be approximated in terms of the
characteristic function of the Gaussian distribution. Since the Edgeworth expansion provides asymptotic correction terms to the Central Limit Theorem up to an order that depends on the number of moments available, it is ideally suited for the analysis of stochastic gravitational wave backgrounds generated by a finite number of astrophysical sources. It is also well-suited for the analysis of signals from cosmological origin in case the deviations from the Gaussian assumption are not too strong. Using a 4th-order Edgeworth expansion, we obtain an explicit expression for the nearly optimal non-Gaussian likelihood statistic. This expression generalizes the standard maximum likelihood detection statistic, which is recovered in the limit of vanishing
third and fourth cumulants of the empirical conditional distribution of the
detector measurement. We use numerical procedures to generate maximum likelihood estimates for the gravitational wave distribution parameters for a set of heavy-tailed distributions and find that the fourth cumulant can be estimated with reasonable precision when the ratio between the signal and the noise variances is larger than 0.01. The use of the non-Gaussian detection statistic comes with no loss of sensitivity when the signal variance is small compared to the noise variance, and involves an efficiency gain when the noise and the signal are of comparable magnitudes.
The rest of the paper is organized as follows. In
section 2, we introduce a detection statistic for a non-Gaussian stochastic
background distribution. 
In section 4, we discuss how the approach can be extended to analyze an individual resolved signal with a non-Gaussian distribution, as opposed to a stochastic background signal emanating from many unresolved astrophysical sources or from cosmological origin. Section 5 contains a conclusion and suggestions for further
research.

\section{\label{sec:section2}Detection Methods for Non-Gaussian Gravitational Wave Stochastic
Backgrounds}

Following DF03, we consider two gravitational wave detectors, assumed to be identical
(which implies that the noise for both detectors is drawn from the same
distribution), as well as coincident and co-aligned (i.e., they have identical
location and arm orientations, which implies that the signal measured by both detectors
is drawn from the same distribution). 

We decompose the measurement output $%
h_{it}$\ for detector $i=1,2$\ at time $t$\ in terms of noise $n_{it}$\
(specific to each detector) versus signal $s_{t}$ (common to both
detectors): $h_{it}=n_{it}+s_{t}.$ Here we assume that the noise in detector 
$1$\ and $2$\ follow uncorrelated Gaussian distributions $\mathcal{N}_{1}$
and $\mathcal{N}_{2}$ with zero mean and standard deviation denoted by $%
\sigma _{1}$\ and $\sigma _{2}$, respectively. We denote by $c_{2}\equiv
\alpha ^{2}$ the variance of the signal distribution $\mathcal{S}$, and by $%
c_{3}$ and $c_{4}$, respectively, the third- and fourth-order cumulants of
the signal distribution. When the signal is Gaussian, we have that $%
c_{3}=c_{4}=0$. Note that we use cap letters for the random variables, e.g., $%
\mathcal{S}$ for the signal, and small letters for their realizations given a given random outcome $\omega $. In other words, we write $\mathcal{S}_{t}(\omega)=s_{t}$, where $\mathcal{S}_{t}$ for $t=1,...,T$  are identical copies of the stationary random variable $\mathcal{S}$, and where $T$ is the number of data points \footnote{In this paper $T$ is not the observation time but the number of data points which is the observation time multiplied by the sampling rate}. 

\subsection{Gram-Charlier and Edgeworth Expansions}

There are three main approaches to statistical problems involving
non-Gaussian distributions. The first approach is the parametric approach,
which consists of assuming a given non-Gaussian distribution and using the
assumed density to derive, subject to analytical tractability, an expression for 
the log likelihood, which can subsequently be used to perform parameter estimation.
To alleviate the concern over specification risk (i.e., the risk that the true unknown distribution differs from the assumed distribution),
one may instead prefer a non parametric approach, where no assumption is made about the unknown distribution, and where sample
information is used to perform parameter estimation. Two main
shortcomings of this approach are the fact that it does not allow for
likelihood maximization, and also the lack of robustness due to the sole
reliance on sample-based information. In what follows, we propose to use a
medium-term, semi-parametric, approach, which allows one to approximate the
unknown density as a transformation of a reference function (typically the
Gaussian density), involving higher-order moments/cumulants of the unknown
distribution. This approach has been heavily used in statistical problems
involving a mild departure from the Gaussian distribution; it is generally
more robust than the non-parametric approach, since the sample information
is only used to generate estimates for the 3rd and 4th cumulants of the
unknown distribution function, and it does not suffer from the specification
risk inherent to the parametric approach. In what follows, we show that
a semi-parametric expansion of the unknown signal density function allows
us to obtain an analytical derivation of the nearly optimal maximum likelihood
detection statistic for non-Gaussian gravitational wave stochastic
backgrounds.

Let $f$ be the density function of the unknown distribution of the
stochastic background signal $S$ which we want to approximate as a function
of the Gaussian density function $\phi$. We first introduce $G$, the \textit{moment
generating function} of $S$:
 \begin{equation}
 G_{s}\left( t\right) =E\left[ e^{tS}\right]
=\int\nolimits_{-\infty }^{\infty }e^{ts}f_{s}\left( x\right) dx
\end{equation}
It is
related to the characteristic distribution $\psi$, i.e., the Fourier transform of the
function $f_{s}$, by $\psi _{s}\left( t\right) =G_{s}\left( it\right)$.

Using the Taylor expansion of the exponential function around 0,
$e^{x}=\sum\limits_{j=0}^{\infty }\frac{x^{j}}{j!}$,
we may obtain the following expression for the characteristic function: 
\begin{equation}
\psi _{s}\left( t\right) =%
E\left[ e^{itS}\right] =\sum\limits_{j=0}^{\infty }\frac{\left( it\right)
^{j}}{j!}E\left( S^{j}\right) \equiv \sum\limits_{j=0}^{\infty }\frac{%
\left( it\right) ^{j}}{j!}\mu _{j},
\end{equation}
where $\mu _{j}$ denotes the $j^{th}$
(non central) moment of the distribution of $S$. From this, we see that $%
j^{th}$ (non central) moment of the distribution is given by the $j^{th}$
derivative of the moment-generating function $G_{s}$ taken at $t=0$ (hence the
name \textit{moment generating function}): 
$\mu _{j}=G_{s}^{\left( j\right)
}\left( 0\right) =\left( -i\right) ^{j}\psi _{s}^{\left( j\right) }\left(
0\right)$. 

We also introduce the \textit{cumulant generating function} $%
g_{s}$
as the logarithm of the characteristic function: 
\begin{equation}
g_{s}\left(
t\right) =\log G_{s}\left( t\right) =\log \sum\limits_{j=1}^{\infty }\frac{\left( it\right) ^{j}}{j!}\mu _{j}.
\end{equation}
A Taylor expansion of the cumulant
generating function $g_{s}$ can be written under the following form: 
\begin{equation}
g_{s}\left( t\right) =g_{s}\left( 0\right) +\sum\limits_{j=1}^{\infty }%
\frac{t^{j}}{j!}g_{s}^{\left( j\right) }\left( 0\right) 
\end{equation}
and we define $c_{j}=g_{s}^{\left( j\right) }\left( 0\right) $
as the $j^{th}$ \textit{%
cumulant} of the random variable $S.$

A moments-to-cumulants
relationship can be obtained by expanding the exponential and equating
coefficients of $t^{j}$ in: 
\begin{equation}
G_{s}\left( t\right) =\exp \left[ g_{s}\left(
t\right) \right] \Longleftrightarrow \sum\limits_{j=0}^{\infty }\frac{t^{j}%
}{j!}\mu _{j}=\exp \left[ \sum\limits_{j=1}^{\infty }\frac{t^{j}}{j!}c_{j}%
\right]. 
\end{equation}
Conversely, a cumulants-to-moments relationship is obtained by
expanding the logarithmic and equating coefficients of $t^{j}$ in
$g_{s}\left( t\right) =\log G_{s}\left( t\right)$. 
Hence we have:%
\begin{eqnarray}
c_{1} &=&g_{s}^{\prime }\left( 0\right) =\mu _{1}=\mu  \label{c1} \\
c_{2} &=&g_{s}^{\prime \prime }\left( 0\right) =\mu _{2}-\mu _{1}^{2}=\sigma
^{2}  \label{c2} \\
c_{3} &=&g_{s}^{(3)}\left( 0\right) =\mu _{3}-3\mu _{2}\mu _{1}+2\mu _{1}^{3}
\label{c3} \\
c_{4} &=&g_{s}^{(4)}\left( 0\right) =\mu _{4}-4\mu _{3}\mu _{1}-3\mu
_{2}^{2}+12\mu _{2}\mu _{1}^{2}-6\mu _{1}^{4}  \label{c4}
\end{eqnarray}
We note in particular that first cumulant is equal to the first moment (the mean), and the second cumulant is equal to the second-centered moment (the variance).\footnote{Cumulants are often simpler than corresponding moments. For example, we have for the Gaussian distribution with mean $\mu $ and variance $\sigma ^{2}$: 
$\mu_{1}=\mu $, $\mu _{2}=\mu ^{2}+\sigma ^{2}$, $\mu _{3}=\mu ^{3}+3 \mu \sigma
^{2}$, $\mu _{4}=\mu ^{4}+6\mu ^{2}\sigma ^{2}+3\sigma ^{4}$, etc. On the other hand, we have $c_{1}=\mu $, $c _{2}=\sigma ^{2}$, and $c_{k}=0$ for $k>2$.}

We may now expand the
unknown non-Gaussian signal distribution $f_{s}$ in terms of a known
distribution with probability density function $\phi $, characteristic
function $\Phi $, and standardized cumulants $\gamma _{j}$. The density $%
\phi $ is generally chosen to be that of the normal distribution. Since we
have for the Gaussian distribution $\gamma _{1}=\mu $, $\gamma _{2}=\sigma ^{2}$, and $\gamma _{j}=0$
for $j>2$, cumulants of order higher than 2 can be regarded as useful measures of deviations
from normality. Using the expression of the characteristic functions for the Gaussian and non-Gaussian distributions in
terms of their cumulants, we have: 
\begin{equation}
\psi _{s}\left( t\right) =\exp \left[
\sum\limits_{j=1}^{\infty }\frac{\left( it\right) ^{j}}{j!}c_{j}\right]
=\exp \left[ \sum\limits_{j=1}^{\infty }\frac{\left( it\right) ^{j}}{j!}%
\left( c_{j}-\gamma _{j}\right) \right] \Phi \left( t\right).
\end{equation}
Given that $\gamma _{j}=$ $c_{j}$ for $j=1,2$, and $\gamma _{j}=0$ for $j>2$, we finally
have: 
\begin{equation}
\psi _{s}\left( t\right) =\exp \left[ \sum\limits_{j=3}^{\infty }%
\frac{\left( it\right) ^{j}}{j!}c_{j}\right] \Phi \left( t\right).
\end{equation}
By the properties of the Fourier transform, $\left( it\right) ^{j}\Phi \left(
t\right) $ is the Fourier transform of $(-1)^{j}D^{j}\phi (x)$, where $D$ is
the differential operator with respect to $x$. From this, we obtain 
\begin{equation}
f_{s}\left( x\right) =\exp \left[ \sum\limits_{j=3}^{\infty }c_{j}\frac{%
\left( -D\right) ^{j}}{j!}\right] \phi \left( x\right).
\end{equation}
Introducing the
Hermite polynomials $H_{j}(\frac{x-\mu }{\sigma })=\left( -1\right)
^{j}\sigma ^{j}\frac{\phi ^{\left( j\right) }\left( x\right) }{\phi \left(
x\right) }$, expanding the exponential and collecting terms according to the
order of the derivatives, we obtain the Gram-Charlier A series, which
is a second-order approximation for a distribution with mean zero and
standard-deviation denoted by $\alpha $:%
\begin{equation}
f_{s}\left( x\right) \simeq \frac{1}{\sqrt{2\pi }\alpha }\exp \left[ -\frac{%
x^{2}}{2\alpha ^{2}}\right] \left[ 1+\frac{c_{3}}{6\alpha ^{3}}H_{3}\left( 
\frac{x}{\alpha }\right) +\frac{c_{4}}{24\alpha ^{4}}H_{4}\left( \frac{x}{%
\alpha }\right) \right]
\end{equation}%
where the 3rd and 4th Hermite polynomials are respectively given by $%
H_{3}(x)=x^{3}-3x$, $H_{4}(x)=x^{4}-6x^{2}+3$. One problem with the
Gram-Charlier A series is that it is not possible to estimate the
error of the expansion. For this reason, another expansion, the Edgeworth
expansion, is generally preferred over the Gram-Charlier expansion. The
Edgeworth expansion is based on the assumption that the unknown signal
distribution is the sum of normalized i.i.d. (non necessarily Gaussian) variables, and provides asymptotic correction terms to the Central Limit Theorem up to an order that depends on the number of moments available. When taken at the fourth-order level the Edgeworth expansion reads as follows (see \cite{feller2008introduction} for the proof, and additional results regarding regarding the convergence rate of the Edgeworth
expansion):%
\begin{equation}
f_{s}\left( x\right) \simeq \phi \left( x\right) \left[ 1+\frac{c_{3}}{%
6\alpha ^{3}}H_{3}\left( \frac{x}{\alpha }\right) +\frac{c_{4}}{24\alpha ^{4}%
}H_{4}\left( \frac{x}{\alpha }\right) +\frac{c_{3}^{2}}{72\alpha ^{6}}%
H_{6}\left( \frac{x}{\alpha }\right) \right] 
\end{equation}%
where the 6th Hermite polynomial is defined as $H_{6}\left( x\right)
=x^{6}-15x^{4}+45x^{2}-15$. We finally have $f_{s}\left( x\right) \simeq
\phi \left( x\right) g\left( x\right) $\ with:%
\begin{eqnarray}
g\left( x\right)  &=&\left( 1+\frac{c_{4}}{8\alpha ^{4}}-\frac{5c_{3}^{2}}{%
24\alpha ^{6}}\right) -\frac{c_{3}}{2\alpha ^{4}}x+\left( \frac{15c_{3}^{2}}{%
24\alpha ^{8}}-\frac{c_{4}}{4\alpha ^{6}}\right) x^{2}  \nonumber \\
&&+\frac{c_{3}}{6\alpha ^{6}}x^{3}+\left( \frac{c_{4}}{24\alpha ^{8}}-\frac{%
5c_{3}^{2}}{24\alpha ^{10}}\right) x^{4}+\frac{c_{3}^{2}}{72\alpha ^{12}}%
x^{6} \\
&\equiv &b_{0}+b_{1}x+b_{2}x^{2}+b_{3}x^{3}+b_{4}x^{4}+b_{6}x^{6}
\end{eqnarray}

We see that the Edgeworth expansion involves one more Hermite polynomial with respect to the Gram-Charlier expansion
while keeping the number of parameters constant. For symmetric
distributions, we have $c_{3}=0$, and the two 4th-order expansion
coincide. In general they differ by the presence of the additional term $%
\frac{c_{3}^{2}}{72\sigma ^{6}}H_{6}\left( \frac{x}{\sigma }\right) $ in the
Edgeworth expansion.\footnote{
One of the problems with both of these expansions is that the resulting
approximate density, while it does integrate to 1, may in general take on negative values. To address this problem \cite{gallant1987semi,gallant1989seminonparametric} suggest to square the polynomial part and then
scale it so as to ensure that the integral of the resulting density is 1. For the parameter values we consider, the problem is not substantial, and we will not apply this transformation.}

\subsection{Nearly Optimal Maximum Likelihood Detection statistic}

We consider here a stochastic background gravitational wave signal having an
unknown distribution with mean zero, variance denoted by $\alpha ^{2}$ or 
equivalently $c_{2}$ and third- and fourth-order cumulants denoted by $c_{3}$ and $c_{4}$, respectively. The standard Bayesian approach for signal detection consists in
finding the value for the unknown parameters, here $(\alpha ,\sigma
_{1},\sigma _{2},c_{3},c_{4})$, where $\sigma_{1}$ and $\sigma_{2}$ denote the standard-deviation of the noise on detectors 1 and 2, so as to minimize the false dismissal
probability at a fixed value of the false alarm probability. This criteria,
known as the Neyman-Pearson criteria, is uniquely defined in terms of the
so-called likelihood ratio: 
\begin{equation}
L=\frac{\left.
p_{h}\right\vert _{X=1}}{\left. p_{h}\right\vert _{X=0}},
\end{equation}
where $\left.
p_{h}\right\vert _{X=1}$\ (respectively, $\left. p_{h}\right\vert _{X=0}$)
is the conditional density for the measurement output if a signal is present
(respectively, absent).  As argued in \cite%
{2003PhRvD..67h2003D} (see page 8, discussion before equation (2.19)), a
natural approximation of the likelihood ratio is the maximum likelihood
detection statistic defined by: 
\begin{equation}
\Lambda _{ML}=\frac{\underset{\alpha ,\sigma _{1},\sigma _{2},c_{3},c_{4}}{%
\max }\int \left. f_{s}\right\vert _{X=1}\left( s\right) \left.
f_{n}\right\vert _{X=1}\left( h-s\right) ds}{\underset{\sigma _{1},\sigma
_{2}}{\max }\left. f_{n}\right\vert _{X=0}\left( h\right) }
\end{equation}

Using the Gaussian assumption for the noise distribution, we obtain after
straightforward manipulations the following expression for the likelihood ratio for the
non-Gaussian distribution: 
\begin{equation}
\Lambda _{ML}^{NG}=\underset{\alpha ,\sigma
_{1},\sigma _{2},c_{3},c_{4}}{\max }\prod\limits_{t=1}^{T}G_{t}\int_{-%
\infty }^{+\infty }\frac{1}{\sigma \sqrt{2\pi }}\exp \left[ -\frac{1}{%
2\sigma ^{2}}\left( s_{t}-\mu _{t}\right) ^{2}\right] g\left( s_{t}\right)
ds_{t}
\end{equation}
 with:
 \begin{equation}
G_{t} \equiv \frac{\sigma }{\alpha }\frac{\overline{\sigma }_{1}\overline{%
\sigma }_{2}}{\sigma _{1}\sigma _{2}}\exp \left[ -\frac{h_{1t}^{2}}{2\sigma
_{1}^{2}}-\frac{h_{2t}^{2}}{2\sigma _{2}^{2}}+1\right] \exp \left[ \frac{1}{2%
}\sigma ^{2}\left( \frac{h_{1t}}{\sigma _{1}^{2}}+\frac{h_{2t}}{\sigma
_{2}^{2}}\right) ^{2}\right] 
\end{equation}
and
 \begin{subequations}
\begin{eqnarray}
\sigma &\equiv &\left( \frac{1}{\alpha ^{2}}+\frac{1}{\sigma _{1}^{2}}+\frac{%
1}{\sigma _{2}^{2}}\right) ^{-\frac{1}{2}} \\
\overline{\sigma }_{i} &\equiv &\sqrt{\frac{1}{T}\sum%
\limits_{t=1}^{T}h_{it}^{2}} \\
\mu _{t} &\equiv &\left( \frac{h_{1t}}{\sigma _{1}^{2}}+\frac{h_{2t}}{\sigma
_{2}^{2}}\right) \sigma ^{2}
\end{eqnarray}
\end{subequations}

Finally, we have: 
\begin{equation}
\Lambda _{ML}=\underset{\alpha ,\sigma _{1},\sigma _{2},c_{3},c_{4}}{\max }%
\prod\limits_{t=1}^{T}G_{t}\left(
I_{0}+I_{1t}+I_{2t}+I_{3t}+I_{4t}+I_{6t}\right)
\end{equation}%
so we are left with the following integrals, which can be obtained from the
first moments of the Gaussian distribution, with the following results: 
 \begin{subequations}
\begin{eqnarray*}
I_{0} &=&b_{0}\int_{-\infty }^{+\infty }\frac{1}{\sigma \sqrt{2\pi }}\exp %
\left[ -\frac{1}{2\sigma ^{2}}\left( s_{t}-\mu _{t}\right) ^{2}\right]
ds_{t}=b_{0} \\
I_{1t} &=&b_{1}\int_{-\infty }^{+\infty }\frac{1}{\sigma \sqrt{2\pi }}%
s_{t}\exp \left[ -\frac{1}{2\sigma ^{2}}\left( s_{t}-\mu _{t}\right) ^{2}%
\right] ds_{t}=b_{1}\mu _{t} \\
I_{2t} &=&b_{2}\int_{-\infty }^{+\infty }\frac{1}{\sigma \sqrt{2\pi }}%
s_{t}^{2}\exp \left[ -\frac{1}{2\sigma ^{2}}\left( s_{t}-\mu _{t}\right) ^{2}%
\right] ds_{t}=b_{2}\left( \mu _{t}^{2}+\sigma ^{2}\right)  \\
I_{3t} &=&b_{3}\int_{-\infty }^{+\infty }\frac{1}{\sigma \sqrt{2\pi }}%
s_{t}^{3}\exp \left[ -\frac{1}{2\sigma ^{2}}\left( s_{t}-\mu _{t}\right) ^{2}%
\right] ds_{t}=b_{3}\left( \mu _{t}^{3}+3\mu _{t}\sigma ^{2}\right)  \\
I_{4t} &=&b_{4}\int_{-\infty }^{+\infty }\frac{1}{\sigma \sqrt{2\pi }}%
s_{t}^{4}\exp \left[ -\frac{1}{2\sigma ^{2}}\left( s_{t}-\mu \right) ^{2}%
\right] ds_{t}=b_{4}\left( \mu _{t}^{4}+6\mu _{t}^{2}\sigma ^{2}+3\sigma
^{4}\right)  \\
I_{6t} &=&b_{6}\int_{-\infty }^{+\infty }\frac{1}{\sigma \sqrt{2\pi }}%
s_{t}^{6}\exp \left[ -\frac{1}{2\sigma ^{2}}\left( s_{t}-\mu _{t}\right) ^{2}%
\right] ds_{t}=b_{6}\left( \mu _{t}^{6}+15\mu _{t}^{4}\sigma ^{2}+45\mu
_{t}^{2}\sigma ^{4}+15\sigma ^{6}\right) 
\end{eqnarray*}
 \end{subequations}
We note that when $c_{3}=c_{4}=0$, that is when the third
and fourth-order cumulant vanish as would be the case for a Gaussian
distribution, then we have $I_{0}=1$, $I_{1}=I_{2}=I_{4}=I_{6}=0,$\ and we
recover the maximum likelihood statistic of the Gaussian case (see DF03 \cite{2003PhRvD..67h2003D} ): 
\begin{equation}
\Lambda _{ML}^{G}=\underset{\alpha ,\sigma
_{1},\sigma _{2}}{\max }\prod\limits_{t=1}^{T}\frac{\sigma }{\alpha }\frac{%
\overline{\sigma }_{1}\overline{\sigma }_{2}}{\sigma _{1}\sigma _{2}}\exp %
\left[ -\frac{h_{1t}^{2}}{2\sigma _{1}^{2}}-\frac{h_{2t}^{2}}{2\sigma
_{2}^{2}}+1\right] \exp \left[ \frac{1}{2}\sigma ^{2}\left( \frac{h_{1t}}{%
\sigma _{1}^{2}}+\frac{h_{2t}}{\sigma _{2}^{2}}\right) ^{2}\right].
\end{equation}

In the Gaussian case, it can be shown that the likelihood ratio admits the
following simple analytical expression (equation 3.13 in DF03 \cite{2003PhRvD..67h2003D}): 
$\Lambda _{ML}^{G}=\left( 1-\frac{\overline{\alpha }^{4}}{\overline{%
\sigma }_{1}^{2}\overline{\sigma }_{2}^{2}}\right) ^{-\frac{T}{2}},
$
 with $%
\overline{\alpha }=\sqrt{\frac{1}{T}\sum\limits_{t=1}^{T}h_{1t}h_{2t}}$.
From a monotonic transformation, this detection statistic is equivalent to
standard the cross-correlation statistic: $\Lambda _{CC}=\sqrt{1-\left(
\Lambda _{ML}^{G}\right) ^{-\frac{2}{T}}}=\frac{\overline{\alpha }^{2}}{%
\overline{\sigma }_{1}\overline{\sigma }_{2}}$. In general, the presence of
the additional terms related to highe-order cumulants implies a correction with respect to the Gaussian case.
This correction makes it impossible to obtain the maximum likelihood estimate in closed-form, but straightforward numerical procedures can be used to maximize the log-likelihood function (see next Section). 

The maximum likelihood estimator is attractive since it is well-known to enjoy a number of desirable properties, including notably consistency and asymptotic efficiency. On the other hand, we now show that one can also use a moment-based method for an analytical estimation of the higher-order cumulants, thus alleviating the need to perform numerical log-likelihood maximization. The moment-based estimate for the variance of the signal coincides with the maximum likelihood estimator, but this correspondence does not extend to higher-order moments and the analytical moment-based estimators for $c_{3}$ and $c_{4}$ do not coincide in general with the the maximum likelihood estimators. In numerical examples, we find that the estimated values are relatively close, but obtain a lower variance for the maximum likelihood (see Table II in Section IV), thus confirming for large sample sizes the superiority (asymptotic efficiency) of the maximum likelihood estimator.

To derive the moment-based estimators, we first write:
\begin{eqnarray*}
\mathbb{E}\left( \mathcal{H}_{1}\mathcal{H}_{2}\right) &=&\mathbb{E}\left[
\left( \mathcal{N}_{1}\mathcal{+S}\right) \left( \mathcal{N}_{2}\mathcal{+S}%
\right) \right] \\
&=&\mathbb{E}\left[ \mathcal{N}_{1}\mathcal{N}_{2}\right] +\mathbb{E}\left[ 
\mathcal{N}_{1}\mathcal{S}\right] +\mathbb{E}\left[ \mathcal{N}_{2}\mathcal{S%
}\right] +\mathbb{E}\left[ \mathcal{S}^{2}\right] \\
&=&\mathbb{E}\left[ \mathcal{N}_{1}\right] \mathbb{E}\left[ \mathcal{N}_{2}%
\right] +\mathbb{E}\left[ \mathcal{N}_{1}\right] \mathbb{E}\left[ \mathcal{S}%
\right] +\mathbb{E}\left[ \mathcal{S}\right] \mathbb{E}\left[ \mathcal{N}_{2}%
\right] +\mathbb{E}\left[ \mathcal{S}^{2}\right] \text{ by independence} \\
&=&\mathbb{E}\left[ \mathcal{S}^{2}\right] =\alpha ^{2}\text{ since all
distributions have zero mean}
\end{eqnarray*}

From this analysis, we obtain that the empirical counterpart for 
$\mathbb{E}\left( \mathcal{H}_{1}\mathcal{H}_{2}\right) $, namely $\frac{1}{T%
}\sum\limits_{t=1}^{T}h_{1t}h_{2t}$ is a natural estimator for the quantity 
$\alpha ^{2}$, an estimator we may call $\widehat{\alpha }^{2}$.
It turns out that it can be shown that this estimator coincides with the
Gaussian maximum likelihood estimator (DF03 \cite{2003PhRvD..67h2003D}).

We also have:
\begin{eqnarray*}
\mathbb{E}\left( \mathcal{H}_{1}\mathcal{H}_{2}^{2}\right) &=&\mathbb{E}%
\left[ \left( \mathcal{N}_{1}\mathcal{+S}\right) \left( \mathcal{N}_{2}%
\mathcal{+S}\right) ^{2}\right] \\
&=&\mathbb{E}\left[ \left( \mathcal{N}_{1}\mathcal{+S}\right) \left( 
\mathcal{N}_{2}^{2}\mathcal{+S}^{2}+2\mathcal{N}_{2}\mathcal{S}\right) %
\right] \\
&=&\mathbb{E}\left[ \mathcal{N}_{1}\mathcal{N}_{2}^{2}\right] +\mathbb{E}%
\left[ \mathcal{N}_{1}\mathcal{S}^{2}\right] +2\mathbb{E}\left[ \mathcal{N}%
_{1}\mathcal{N}_{2}\mathcal{S}\right] +\mathbb{E}\left[ \mathcal{SN}_{2}^{2}%
\right] +\mathbb{E}\left[ \mathcal{S}^{3}\right] +2\mathbb{E}\left[ \mathcal{%
S}^{2}\mathcal{N}_{2}\right] \\
&=&\mathbb{E}\left[ \mathcal{S}^{3}\right] =\mu _{3}\text{ since all other
terms are zero }
\end{eqnarray*}

We thus obtain that the empirical counterpart for 
$\mathbb{E}\left( \mathcal{H}_{1}\mathcal{H}_{2}^{2}\right) $, namely $\frac{%
1}{T} \sum\limits_{t=1}^{T}h_{1t}h_{2t}^{2}$ is a natural estimator for the
quantity $\mu _{3}$, an estimator we may call $\widehat{\mu }_{3}$%
. Of course, we would also have that $\mathbb{E}\left( \mathcal{H}_{1}^{2}%
\mathcal{H}_{2}\right) =\mathbb{E}\left[ \mathcal{S}^{3}\right] =\mu _{3}$,
so that we can in the end propose the following estimator for $\mu _{3}$:%
\begin{equation}
\widehat{\mu }_{3}=\frac{1}{2}\left[ \frac{1}{T}\sum%
\limits_{t=1}^{T}h_{1t}h_{2t}^{2}+\frac{1}{T}\sum%
\limits_{t=1}^{T}h_{1t}^{2}h_{2t}\right]
\end{equation}

Finally, we have that:
\begin{eqnarray*}
\mathbb{E}\left( \mathcal{H}_{1}^{2}\mathcal{H}_{2}^{2}\right) &=&\mathbb{E}%
\left[ \left( \mathcal{N}_{1}\mathcal{+S}\right) ^{2}\left( \mathcal{N}_{2}%
\mathcal{+S}\right) ^{2}\right] \\
&=&\mathbb{E}\left[ \left( \mathcal{N}_{1}^{2}\mathcal{+S}^{2}+2\mathcal{N}%
_{1}\mathcal{S}\right) \left( \mathcal{N}_{2}^{2}\mathcal{+S}^{2}+2\mathcal{N%
}_{2}\mathcal{S}\right) \right] \\
&=&\mathbb{E}\left[ \mathcal{S}^{4}\right] +\mathbb{E}\left[ \mathcal{N}%
_{1}^{2}\right] \mathbb{E}\left[ \mathcal{S}^{2}\right] +\mathbb{E}\left[ 
\mathcal{N}_{2}^{2}\right] \mathbb{E}\left[ \mathcal{S}^{2}\right] +\mathbb{E%
}\left[ \mathcal{N}_{1}^{2}\right] \mathbb{E}\left[ \mathcal{N}_{1}^{2}%
\right] \text{ all other terms being zero} \\
&=&\mu _{4}+\alpha ^{2}\left( \sigma _{1}^{2}+\sigma _{2}^{2}\right) +\sigma
_{1}^{2}\sigma _{2}^{2}
\end{eqnarray*}

If we assume that $\sigma _{1}$ and $\sigma _{2}$ are known, then we obtain
the following natural estimator for $\mu _{4}$:
\begin{equation}
\widehat{\mu }_{4}=\frac{1}{T}\sum\limits_{t=1}^{T}h_{1t}^{2}h_{2t}^{2}-%
\left( \sigma _{1}^{2}+\sigma _{2}^{2}\right) \left( \frac{1}{T}%
\sum\limits_{t=1}^{T}h_{1t}h_{2t}\right) -\sigma _{1}^{2}\sigma _{2}^{2}
\end{equation}

In general, the parameters $\sigma _{1}$ and $\sigma _{2}$ are not known. The relationship 
\begin{equation}
\mathbb{E}\left[ \mathcal{H}_{i}^{2}\right] \mathcal{=}\mathbb{E}\left[
\left( \mathcal{N}_{i}\mathcal{+S}\right) ^{2}\right] =\mathbb{E}\left[ 
\mathcal{N}_{i}^{2}\right] +\mathbb{E}\left[ \mathcal{S}^{2}\right]
\end{equation}%
suggests that they can be estimated as follows: 
\begin{equation}
\widehat{\sigma }_{i}^{2}=\overline{\sigma }_{i}^{2}-\widehat{\alpha }^{2}
\end{equation}%
where we recall that: 
\begin{equation}
\overline{\sigma }_{i}^{2}=\frac{1}{T}\sum\limits_{t=1}^{T}h_{it}^{2}
\end{equation}

If we are instead interested in estimates for the first 4 cumulants, we have
that (keeping in mind that the mean of both the signal and the noise distributions is zero):
\begin{eqnarray}
\widehat{\alpha }^{2} &=&\frac{1}{T}\sum\limits_{t=1}^{T}h_{1t}h_{2t} \\
\widehat{c}_{3} &=&\widehat{\mu }_{3} \\
\widehat{c}_{4} &=&\widehat{\mu }_{4}-3\widehat{\mu }_{2}=\widehat{\mu }%
_{4}-3\widehat{\alpha }^{4}
\end{eqnarray}

In the limit of vanishing noise, that is when $\sigma _{1}=\sigma _{2}=0$,
it is straightforward to note that: 
\begin{eqnarray}
&&\widehat{\alpha }\underset{T\rightarrow \infty }{\longrightarrow }\alpha \\
&&\widehat{c}_{3}\underset{T\rightarrow \infty }{\longrightarrow }c_{3} \\
&&\widehat{c}_{4}\underset{T\rightarrow \infty }{\longrightarrow }c_{4}
\end{eqnarray}

\subsection{Implications for SGWB Signal Detection}


For a Gaussian signal, the cross-correlation detection statistic, which can be obtained as a monotonic transformation of the likelihood ratio, is optimal in the sense of minimizing the false dismissal probability at a fixed value of the false alarm probability under restrictive assumptions (DF03 \cite{2003PhRvD..67h2003D}). In the general non-Gaussian case, the  cross-correlation detection statistic may not be optimal, and it is unclear how one could derive the optimal detection statistic in the absence of anlytical expressions for the likelihood ratio. 

What we show, however, is that the application of the standard (a priori sub-optimal) cross-correlation detection statistic allows for a lower probability of a false dismissal for a given detection threshold value, or equivalently to a lower detection threshold level for a given probability of a false dismissal, when the deviation from the Gaussian approximation is explicitly accounted for compared to a situation where the signal distribution is supposed to be Gaussian. This improvement is small when the signal is strongly dominated by the noise, but it is substantial when the signal and the noise are of similar magnitudes. Since the non-Gaussian detection methodology nests the Gaussian methodology as a specific case when the third and fourth cumulants of the signal distribution are zero, it should be noted that the use of this approach can be recommended even when there is uncertainty regarding whether the signal distribution is Gaussian or not.

Our formal argument is based on the analysis of the asymptotic
distribution of the detection statistic in the Gaussian versus non-Gaussian case. We first define the detection statistic $DS$ as being given by:%
\begin{equation}
DS=\sum\limits_{t=1}^{T}\mathcal{H}_{1t}\mathcal{H}_{2t}
\end{equation}%
where $\mathcal{H}_{it}=\mathcal{N}_{it}+\mathcal{S}_{t}$, and where $%
\mathcal{N}_{it}$ and $\mathcal{S}_{t}$, for $1\leq t\leq T,$ are $T$
independent copies of the random variables $\mathcal{N}_{i}$ and $\mathcal{S}
$, respectively.

We have:%
\begin{equation}
DS=\sum\limits_{t=1}^{T}\mathcal{N}_{1t}\mathcal{N}_{2t}+\sum%
\limits_{t=1}^{T}\mathcal{N}_{1t}\mathcal{S}_{t}+\sum\limits_{t=1}^{T}%
\mathcal{N}_{2t}\mathcal{S}_{t}+\sum\limits_{t=1}^{T}\mathcal{S}_{t}^{2}
\end{equation}

A signal is presumed to be detected when the detection statistic $DS$
exceeds a given detection threshold $DT$: 
\begin{equation}
DS>DT
\end{equation}

We typically select the detection threshold $DT$ such that $pfa=x\%$,
where $x\%$ is a given confidence level, and where the probability of a
false alarm is given by:%
\begin{equation}
pfa =\Pr \left( \left. DS>DT\right\vert \mathcal{H}_{it}=\mathcal{N}%
_{it}\right) 
\end{equation}

Obviously, we have that $pfa$ is independent of the signal distribution, so we
turn to the more interesting term, which is the probability of a false
detection $pfd$ given by :
\begin{equation}
pfd =\Pr \left( \left. DS<DT\right\vert \mathcal{H}_{it}=\mathcal{N}_{it}+%
\mathcal{S}_{t}\right) 
\end{equation}
By the strong law of large number, we have that (where $a.s.$
stands for \textit{almost surely}):
\begin{eqnarray}
\sum\limits_{t=1}^{T}\mathcal{N}_{1t}\mathcal{N}_{2t}\overset{a.s.}{%
\underset{T\rightarrow \infty }{\longrightarrow }}\mathbb{E}\left( \mathcal{N%
}_{1}\mathcal{N}_{2}\right)  &=&\mathbb{E}\left( \mathcal{N}_{1}\right) 
\mathbb{E}\left( \mathcal{N}_{2}\right) =0 \\
\sum\limits_{t=1}^{T}\mathcal{N}_{1t}\mathcal{S}\overset{a.s.}{\underset{%
T\rightarrow \infty }{\longrightarrow }}\mathbb{E}\left( \mathcal{N}_{1}%
\mathcal{S}\right)  &=&\mathbb{E}\left( \mathcal{N}_{1}\right) \mathbb{E}%
\left( \mathcal{S}\right) =0 \\
\sum\limits_{t=1}^{T}\mathcal{N}_{2t}\mathcal{S}\overset{a.s.}{\underset{%
T\rightarrow \infty }{\longrightarrow }}\mathbb{E}\left( \mathcal{N}_{2}%
\mathcal{S}\right)  &=&\mathbb{E}\left( \mathcal{N}_{2}\right) \mathbb{E}%
\left( \mathcal{S}\right) =0
\end{eqnarray}

For finite observation times, the contribution of these three terms to the variance of the detection statistic will not vanish, and the variance of the 
\textit{exact} detection statistic $\sum\limits_{t=1}^{T}\mathcal{N}_{1t}%
\mathcal{N}_{2t}+\sum\limits_{t=1}^{T}\mathcal{N}_{1t}\mathcal{S}%
_{t}+\sum\limits_{t=1}^{T}\mathcal{N}_{2t}\mathcal{S}_{t}+\sum%
\limits_{t=1}^{T}\mathcal{S}_{t}^{2}$ will contain contributions from the 4 terms. In what follows, we first assume that the noise is small and focus on the following approximation when the signal is present:%
\begin{equation}
DS=\sum\limits_{t=1}^{T}\mathcal{H}_{1t}\mathcal{H}_{2t}\simeq
\sum\limits_{t=1}^{T}\mathcal{S}_{t}^{2}
\end{equation}%

In this context, we have:
\begin{eqnarray*}
pfd &=&\Pr \left( \left. DS<DT\right\vert H_{it}=\mathcal{N}_{it}+\mathcal{S}%
_{t}\right)  \\
&\simeq &\Pr \left( \sum\limits_{t=1}^{T}\mathcal{S}_{t}^{2}<DT\right) 
\end{eqnarray*}

When the signal is Gaussian, $\sum\limits_{t=1}^{T}\mathcal{S}_{t}^{2}$
follows, by definition, a chi-squared distribution with $T$ degrees of
freedom.\footnote{
When $T\rightarrow \infty $, we know that the chi-squared distribution with $%
T$ degrees of freedom converges towards a Gaussian distribution. For
practical purposes, for $T>50$, the distribution is sufficiently close to a
normal distribution for the difference to be ignored \cite{box1978statistics}.}
On the other hand, when the signal is not Gaussian, it is not
obvious to see what the distribution of the approximate detection statistic 
$\sum\limits_{t=1}^{T}\mathcal{S}_{t}^{2}$ is \textit{for a finite }$T$,
except for very particular choices of non-Gaussian distributions. In
principle, one could use an Edgeworth expansion in order to approximate the
distribution of the detection statistic for each given non-Gaussian signal
distribution. Fortunately, a central limit theorem exists for the sample
variance, which allows us to obtain the \textit{asymptotic} distribution of
the detection statistic as $T$ grows to infinity for \textit{any}
underlying non-Gaussian signal distribution.

Formally, let $S_{1},S_{2},...,S_{T}\ $be $T$ i.i.d. copies of the SBGW signal, each
of them with mean $0$, variance $c_{2}\equiv \alpha ^{2}$, and third and
fourth-order cumulants $c_{3}$ and $c_{4}$. Then, it can be shown that: 
\begin{equation}
\Pr \left( \sqrt{T}\left( \frac{1}{T}\sum\limits_{t=1}^{T}\mathcal{S}_{t}^{2}-\alpha
^{2}\right) <x\right) \underset{T\rightarrow \infty }{\longrightarrow }\Pr
\left( U<x\right) 
\end{equation}%
where $U$ is a Gaussian distribution with mean zero and variance $\sigma
_{U}^{2}=c_{4}+2\alpha ^{4}$.

The proof for this result is straightforward and follows from applying the central limit theorem to squared
signal distributions $S^{2}$ (see for example \cite{omey2008central}).

In other words, we obtain that the detection statistic asymptotically
converges to a Gaussian distribution, with a variance given by a function of
the second and fourth cumulants. More precisely, we have $%
\sum\limits_{t=1}^{T}\mathcal{S}_{t}^{2}=T\widehat{V}_{T}\underset{T\rightarrow
\infty }{\sim }\mathcal{N}\left( T\alpha ^{2},T\left( c_{4}+2\alpha
^{4}\right) \right) $. Note that the \textit{approximate} detection
statistic $\sum\limits_{t=1}^{T}\mathcal{S}_{t}^{2}$ is closely related \
to the \textit{sample} variance of the signal distribution, which is denoted
by $\widehat{V}_{T}=\frac{1}{T}\sum\limits_{t=1}^{T}\mathcal{S}_{t}^{2}$. $%
\widehat{V}_{T}$ admits the following asymptotic distribution $\widehat{V}%
_{T}\underset{T\rightarrow \infty }{\sim }\mathcal{N}\left( \alpha ^{2},%
\frac{1}{T}\left( c_{4}+2\alpha ^{4}\right) \right) $, which shows that it
is an asymptotically unbiased estimator for the signal variance $\alpha ^{2}$%
. The advantage of using $\widehat{V}_{T}$, as opposed to $%
\sum\limits_{t=1}^{T}\mathcal{S}_{t}^{2}$, as a detection statistic is
that the expectation of the former random variable does not depend on $T$.

When the signal is normally distributed, we have that $c_{4}=0$, and
therefore $\sigma _{U,G}^{2}=2\alpha ^{4}$, which is a standard result
regarding the asymptotic distribution of the sample variance in the Gaussian
case. We find that the variance of the asymptotic distribution of the signal
detection statistic for non-Gaussian signal distribution $\sigma
_{U,NG}^{2}=c_{4}+2\alpha ^{4}$ is always greater than the variance of the
Gaussian detection $\sigma _{U,G}^{2}=2\alpha ^{4}$. 

In practice, the detection threshold $DT$ is chosen with sufficiently low
value to correspond to high confidence levels (that
is, $pfa$ and $pfd$ probabilities of 5\% or 10\%). In this context, because of the fatter tails of the distribution of the detection statistic in the non-Gaussian case, the detection threshold corresponding to a given $pfd$ will be lower in the non-Gaussian case (see Figure 1).   

One implication of these findings is that if an observer wrongly uses the
assumption that the signal is Gaussian ($c_{4}=0$) while the signal is truly
non-Gaussian ($c_{4}>0$), then for a given confidence level, the observer using a non-Gaussian
methodology will be using a lower
detection threshold, which in turn will allow
for the detection of fainter signals.

In the realistic case when the noise is present and not small compared to the
noise, on the other hand, we have to account for the contribution of the noise to the variance of the detection statistic. In this case, it can be shown that the detection statistic $\widehat{V}_{T}$ is
asymptotically normally distributed, with mean $\alpha ^{2}$ and variance
given by $\frac{1}{T}\left( \sigma _{1}^{2}\sigma _{2}^{2}+\sigma
_{1}^{2}\alpha ^{2}+\sigma _{2}^{2}\alpha ^{2}+2\alpha ^{4}+c_{4}\right) $.
We therefore find that the variance of the detection statistic when the
non-Gaussianity of the signal is taken into account ($c_{4}>0$) is always
strictly greater than when the signal is assumed to be Gaussian ($c_{4}=0$).
While this correction leads in principle to a sensitivity gain as explained
before, the magnitude of the gain is expected to be small if the noise is
several orders of magnitude larger than the signal, that is when $\frac{1}{T}%
\left( \sigma _{1}^{2}\sigma _{2}^{2}+\sigma _{1}^{2}\alpha ^{2}+\sigma
_{2}^{2}\alpha ^{2}+2\alpha ^{4}+c_{4}\right) \simeq \frac{\sigma
_{1}^{2}\sigma _{2}^{2}}{T}$. 

\begin{figure}
\includegraphics[width=\textwidth]{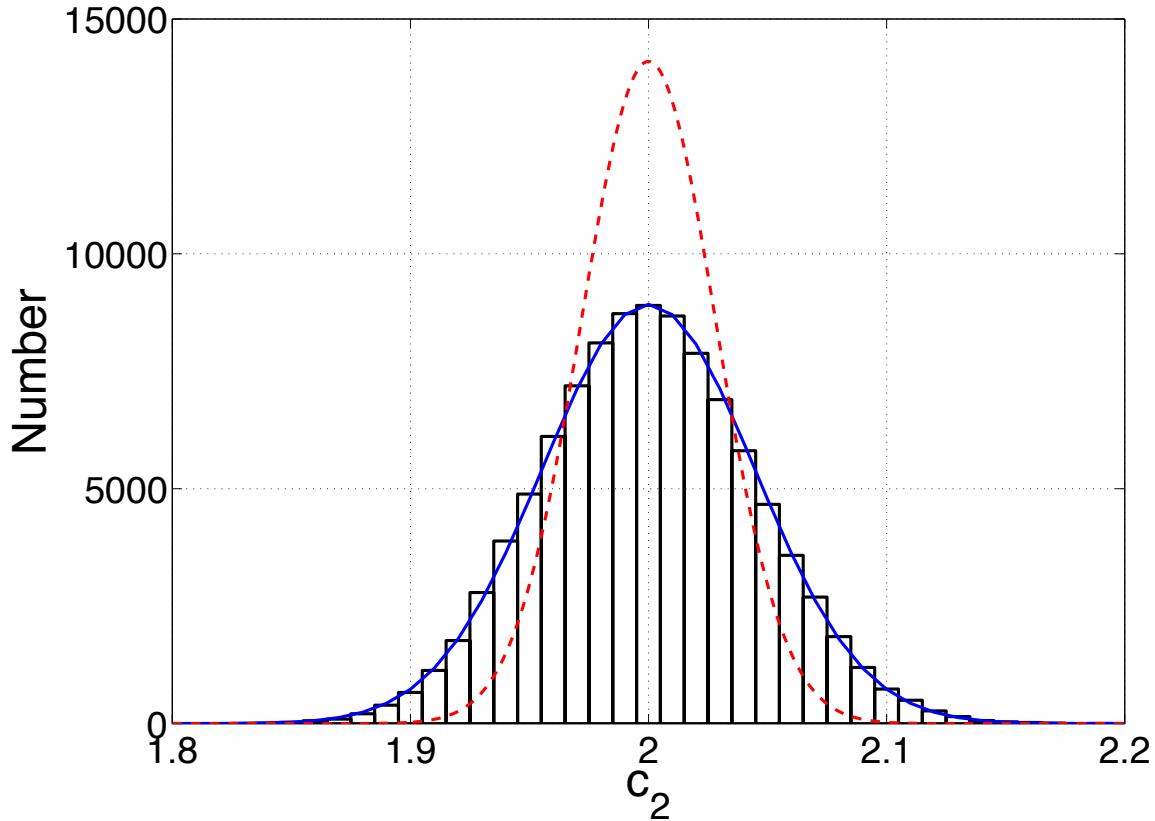}
 \caption{\label{fig3}  Histogram of the estimator for $c_2$ over $10^5$ realizations for the Laplace distribution and for a number of data point $T=10^4$. We assume here that the variance of the signal is 1 and the fourth cumulant is 3. The continuous line shows the non-Gaussian statistic density (a Gaussian centered around 2 with variance $\sigma
_{U,NG}^{2}=c_{4}+2\alpha ^{4}$  and the dashed line the Gaussian statistic density (a Gaussian centered in 2 with variance $\sigma
_{U,G}^{2}=c_{4}$.}
\end{figure}

\section{Numerical Illustrations}

\subsection{Edgeworth expansions of usual distributions}
\begin{table}
\caption{\label{tab:Table1} Test distributions used in this paper, their probability density and cumulants. Note that the $B_{j}$ are the so-called Bernoulli numbers ($B_{0}$=1, $B_{1}$=1/2, $B_{2}$=1/6, $B_{3}$=0, $B_{4}$=-1/30). The $K_{1}$ function is the Bessel function of the first kind.}
\begin{ruledtabular}
\begin{tabular}{lll}
PDF name & Probability density & Cumulants of order $j \left( c_{j}\right) $ \\
\hline
Laplace & $\frac{a}{2}\exp \left( -a\left\vert x\right\vert \right) 
$ & $ c_{2j} =\frac{2\left( 2j-1\right) !}{a^{2j}}$ \\ 
Hypersecant & $\frac{1}{2a}\sec h \left( \frac{\pi }{2a}%
x\right) $ & $c_{2j} =\left( -1\right) ^{j+1}\left(
2^{2j}-1\right) 2^{2j-1} a^{2j} \frac{B_{2j}}{j}$ \\ 
Logistic & $\frac{\exp \left( -\frac{x}{a}\right) }{a\left( 1+\exp
\left( -\frac{x}{a}\right) \right) ^{2}}$ & $c_{2j}=\left( -1\right) ^{j-1}%
\frac{\left( 2a\pi \right) ^{2j}}{2j}B_{2j}$ \\
NIG & $\frac{\delta a\exp \left( \delta a\right) }{\pi \sqrt{\delta
^{2}+x^{2}}}K_{1}\left( a\sqrt{\delta ^{2}+x^{2}}\right)$  & $%
c_{2}=\frac{\delta }{a}$ ; $c_{4}=\frac{3\delta }{a^{3}}$ \\
\end{tabular}
\end{ruledtabular}
\end{table}

We now present a series of numerical illustrations showing how the methodology introduced in this paper can be applied to estimate not only the variance but also the fourth cumulant of the signal distribution. We first consider a list of 5 standard distributions, including the Gaussian distribution as well as 4 non-Gaussian
distributions, namely the Laplace, hypersecant, logistic and normal inverse Gaussian distributions. These distributions are characterized in parametric form by their densities, with
cumulants expressed as a function of the parameters of the density (see Table ~\ref{tab:Table1}).
 The deviation from
the Gaussian is measured in terms of the $c_{3}$ and $c_{4}$ parameters,
which are zero for the Gaussian, and non-zero for non-Gaussian
distributions. It should be noted that all distributions we analyze are symmetric, which implies that the parameter $c_{3}=0$. Beside, we choose the parameter values so that all distributions have a unit variance. 
More specifically, we make the following parametric choices.  
\begin{enumerate}
\item{\textbf{Gaussian distribution}}, we take
the parameter $a =1$, so that we have $c_{2}=1,$ $c_{3}=0$, $c_{4}=0$.
\item{\textbf{Laplace distribution}}, we take the parameter $a=1,$ so that we have $%
c_{2}=1,$ $c_{3}=0$, $c_{4}=12$. 
\item{\textbf{Hypersecant distribution},} we have
that $c_{2}=a^{2},$ $c_{3}=0$ and $c_{4}=2a^{4}$; taking $a=1,$ we have that 
$c_{2}=1,$ $c_{3}=0$ and $c_{4}=2.$ 
\item{\textbf{Logistic distribution}}, we have
that $c_{2}=\frac{a^{2}\pi ^{2}}{3},$ $c_{3}=0$ and $c_{4}=\frac{2a^{4}\pi
^{4}}{15}$; taking $a=\frac{\sqrt{3}}{\pi },$ we have that $c_{2}=1,$ $%
c_{3}=0$ and $c_{4}=18/15=1.2.$ 
\item{\textbf{Normal Inverse Gaussian
distribution}}, we can take $a=${\small \ }$\delta =1$, so that we have $%
c_{2}=1,$ $c_{3}=0$, $c_{4}=3$. 
\end{enumerate}
For each non-Gaussian distribution in the
table, we plot on the same graph for the chosen parameter values the
exact density function of the signal $f_{s}\left( x\right) $\ as well as the
approximate density function, where the approximation is given the
Edgeworth expansion $f_{s}^{E}\left( x\right) $. 
\footnote{The Gram-Charlier expansion would exactly coincide with the Edgeworth expansion for these symmetric distributions.}
In addition to the quality
of fit that can be visually assessed from the analysis of the graph, we also
compute a quantitative measure of the approximation error AE as the quadratic
distance between the exact and approximate density using the Edgeworth
expansion:%
\begin{equation}
AE^{E}=\left( \int\nolimits_{-\infty }^{+\infty }\left( f_{s}^{E}\left(
x\right) -f_{s}\left( x\right) \right) ^{2}dx\right) ^{1/2}
\end{equation}
For comparison purposes, we also report the approximation error with a simple
Gaussian approximation, denoted by $AE^{G}$.

\begin{figure}
\includegraphics[width=0.49\textwidth]{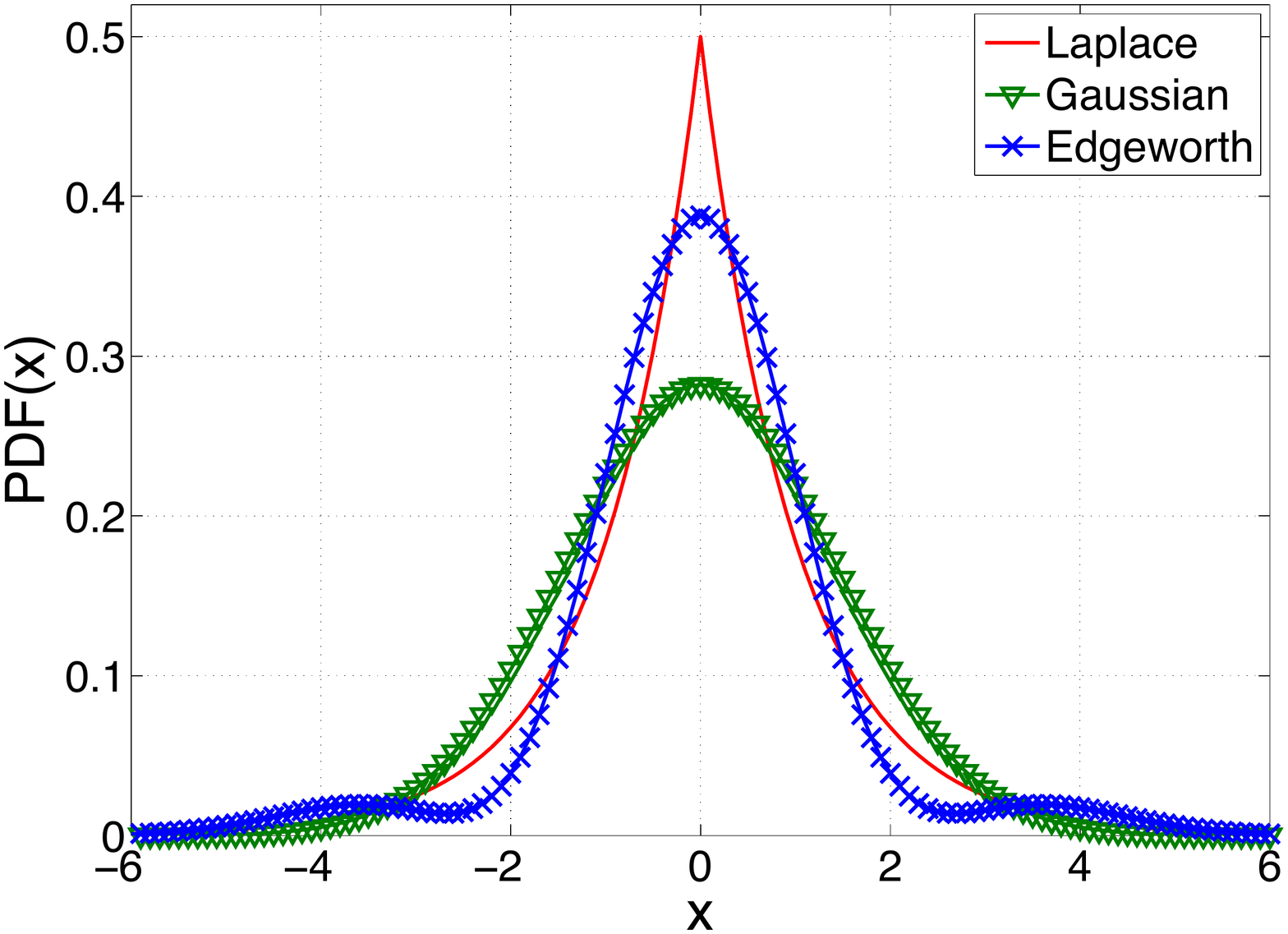}
\includegraphics[width=0.49\textwidth]{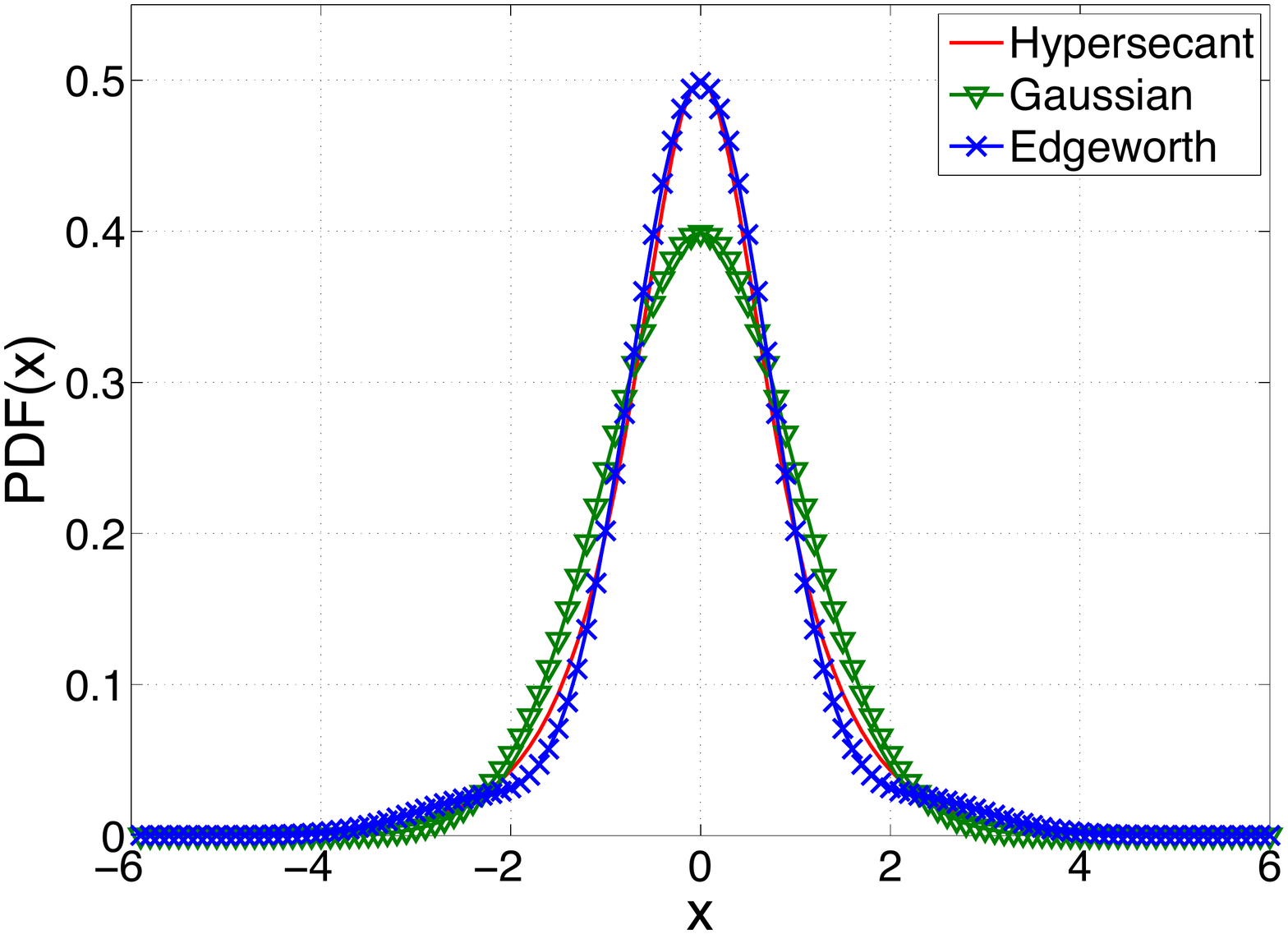}
\includegraphics[width=0.49\textwidth]{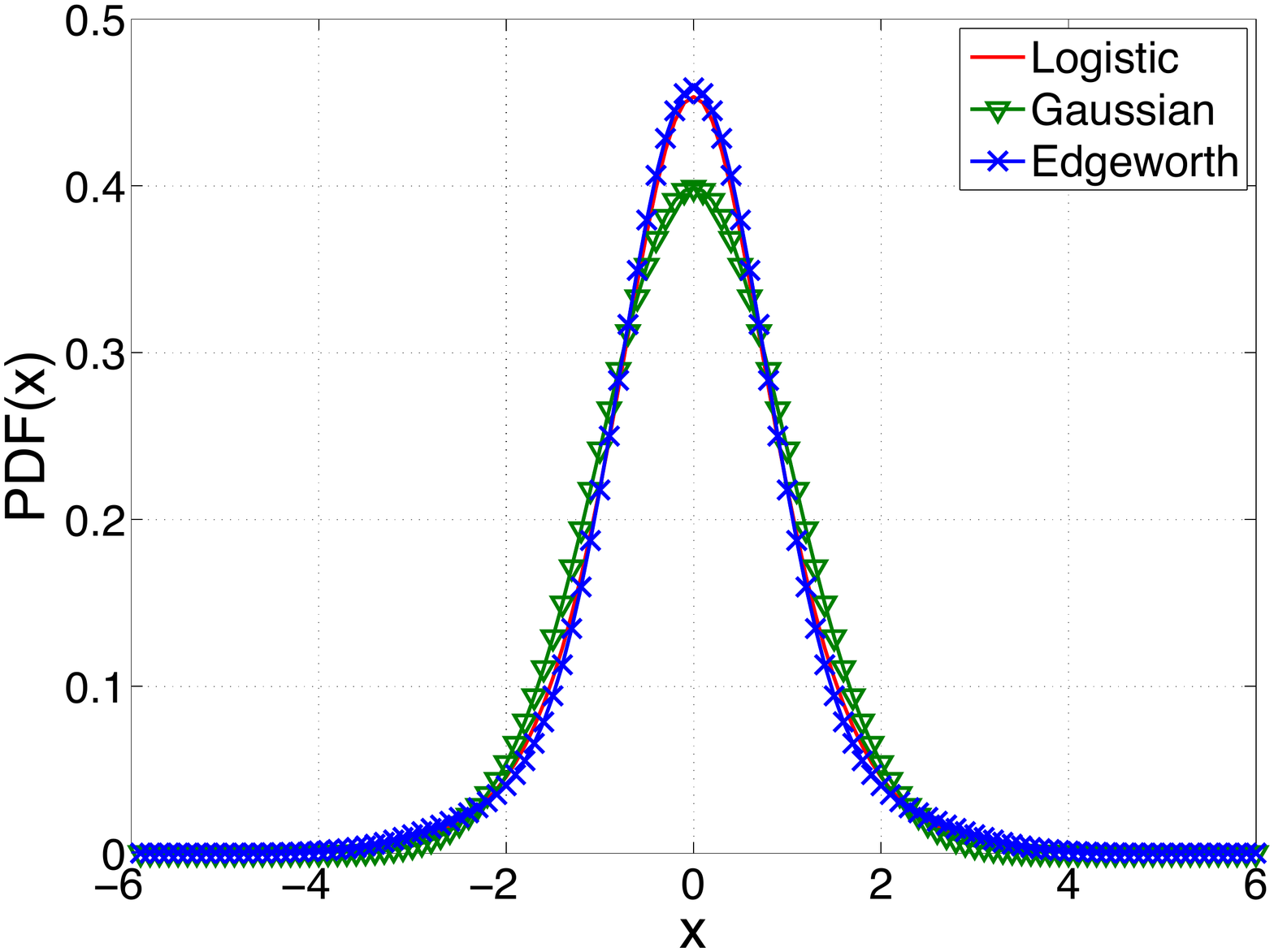}
\includegraphics[width=0.49\textwidth]{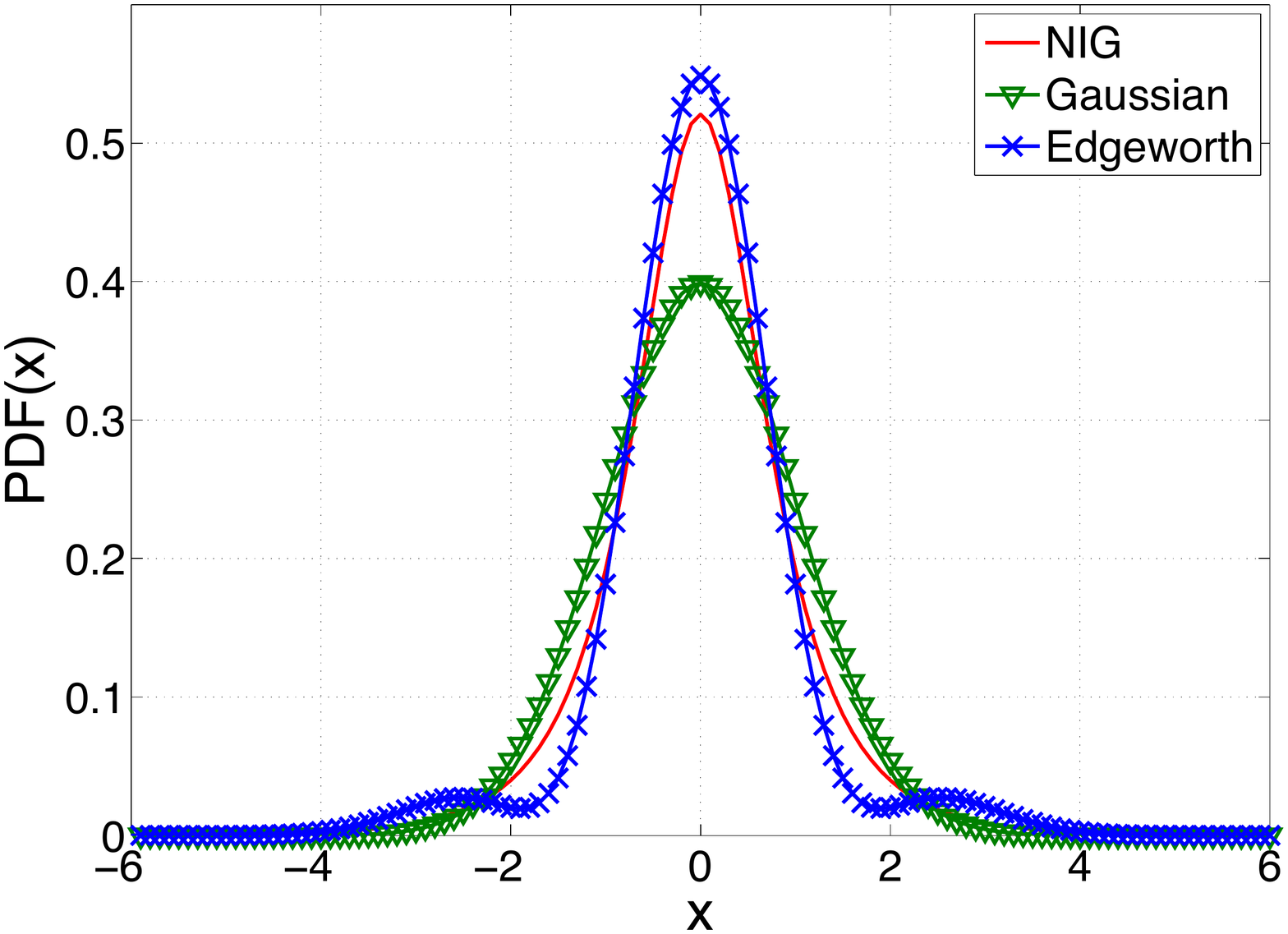}
 \caption{\label{fig1} Edgeworth (crosses) and Gaussian (triangles) approximations for the Laplace (top left), Hypersecant (top right), Logistic (bottom left) and NIG (bottom right) distributions (continuous line). The approximation errors are respectively $AE^{E}=0.085$ and $AE^{G}=0.15$ (Laplace), $AE^{E}=0.03$ and $AE^{G}=0.085$ (Logistic), $AE^{E}=0.015$ and $AE^{G}=0.055$ (Hypersecant), $AE^{E}=0.0046$ and $AE^{G}=0.013$ (NIG). This gives an improvement of $\frac{AE^{G}}{AE^{E}}$ of 1.8 (Laplace), 2.8 (Logistic), 3.6 (Hypersecant) and 2.8 (NIG)
 .}
\end{figure}

The Edgeworth expansions appear to be a better fit
compared to the Gaussian approximation for all non-Gaussian distributions
that we consider, as can be seen from a simple visual inspection, and also
more formally from the fact that approximation errors AE are from 2 to 4
times lower with the non-parametric expansions compared to what is obtained
with the Gaussian approximation. The best fit is given by the NIG distribution with $\frac{AE^{G}}{AE^{E}}=1.8$ and the worse fit by the Laplace distribution with $\frac{AE^{G}}{AE^{E}}=3.6$.

\subsection{Monte Carlo simulations and predictions}

\begin{figure}
\includegraphics[width=0.49\textwidth]{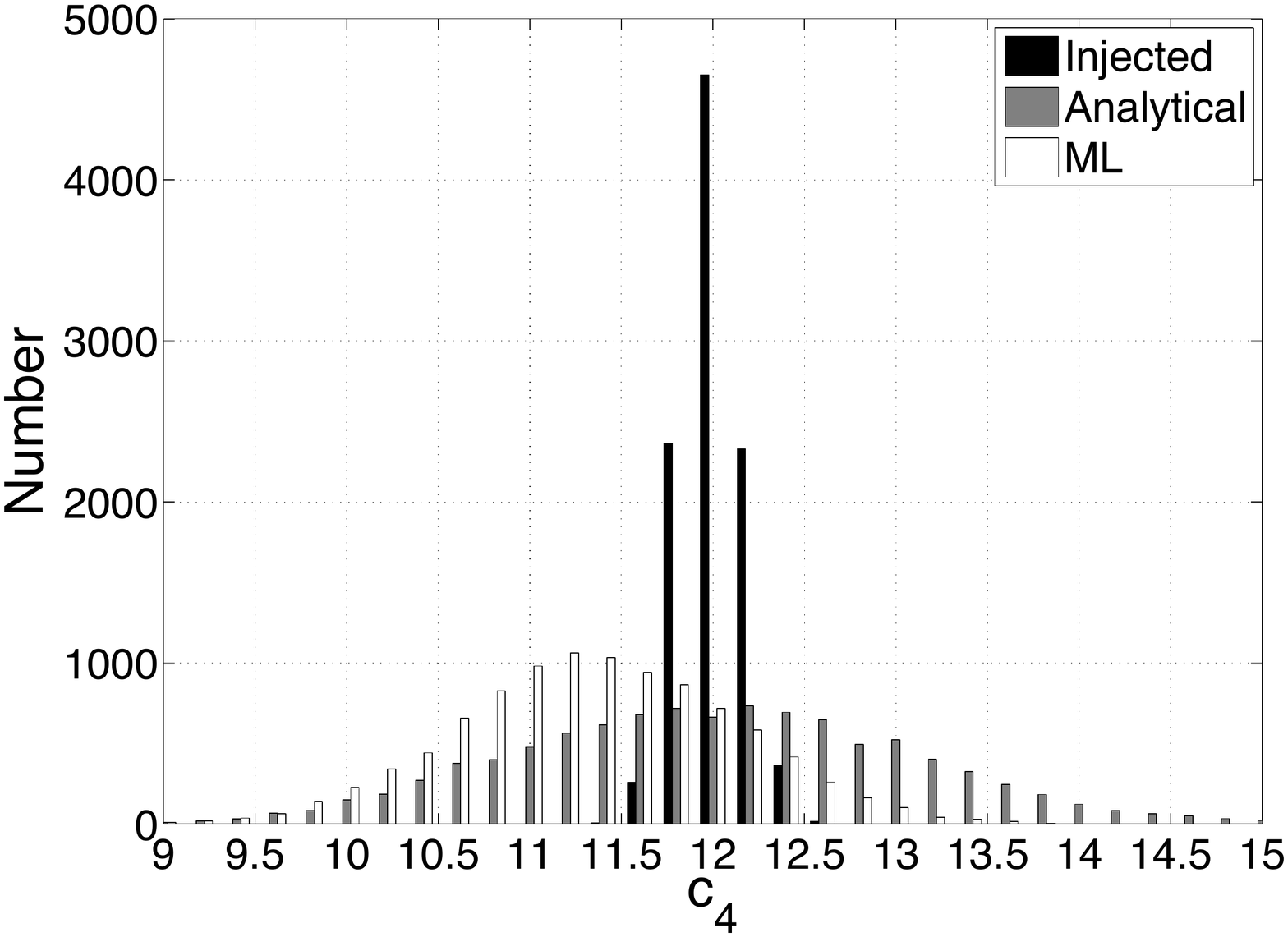}
\includegraphics[width=0.49\textwidth]{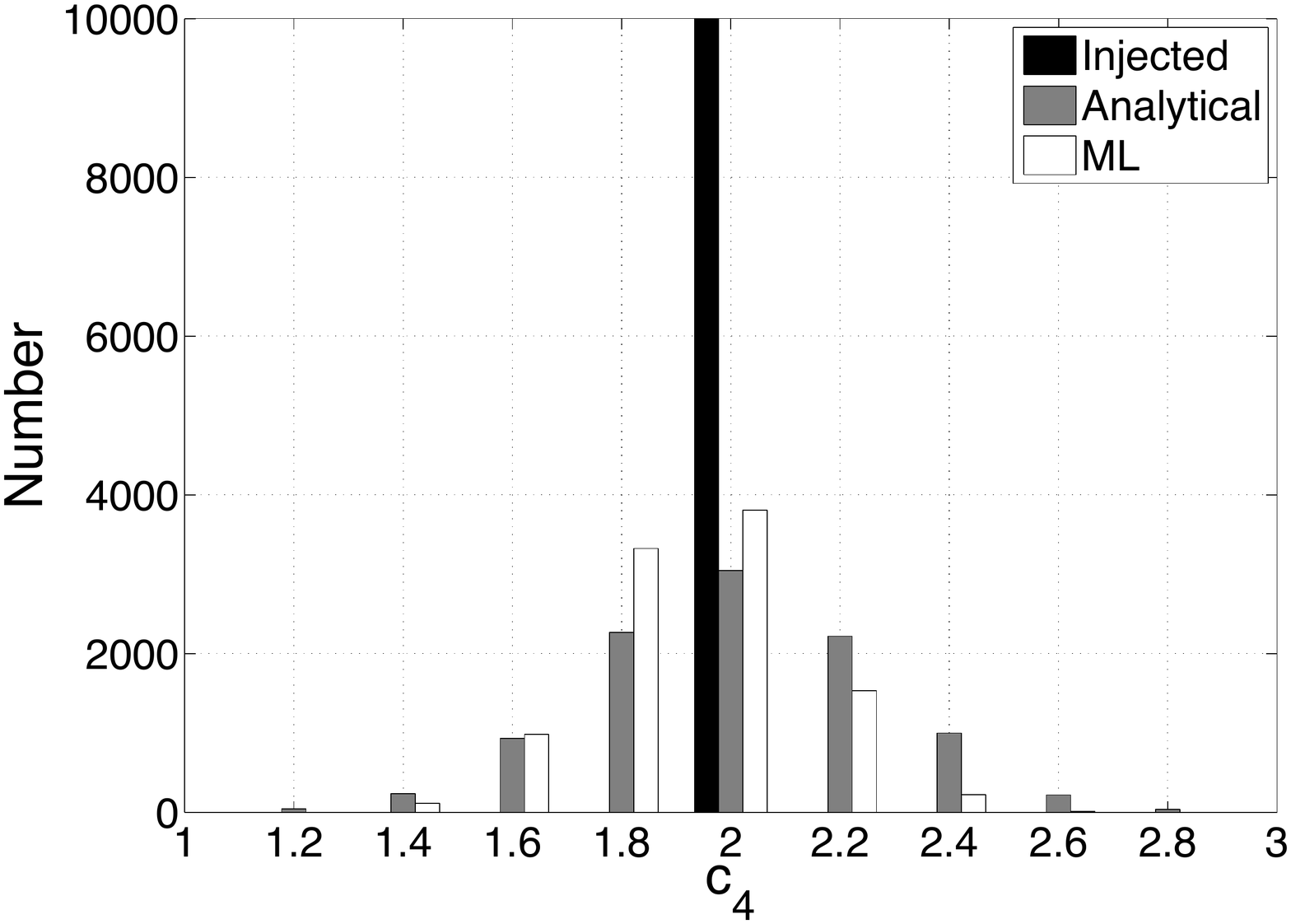}\\
\includegraphics[width=0.49\textwidth]{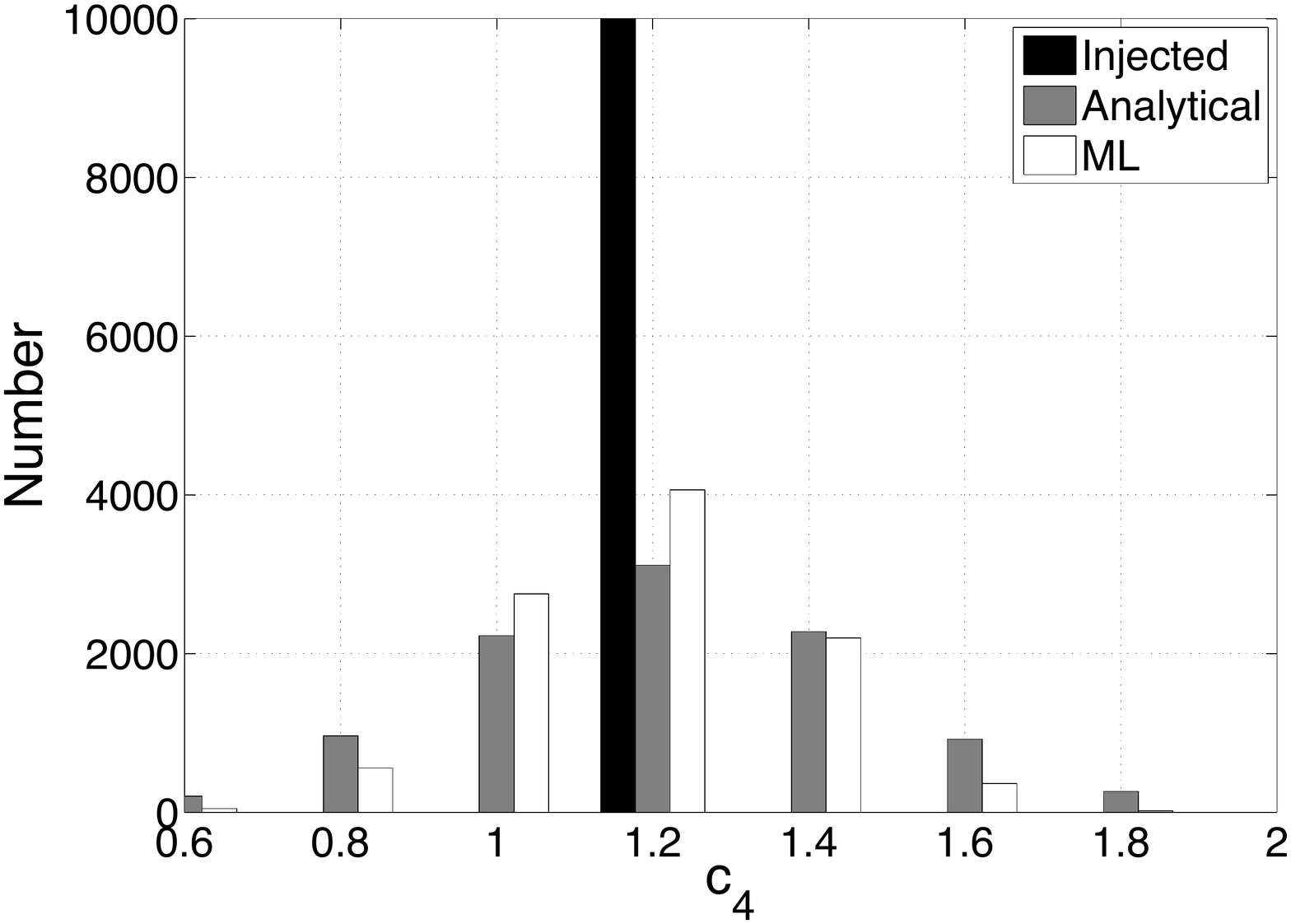}
\includegraphics[width=0.49\textwidth]{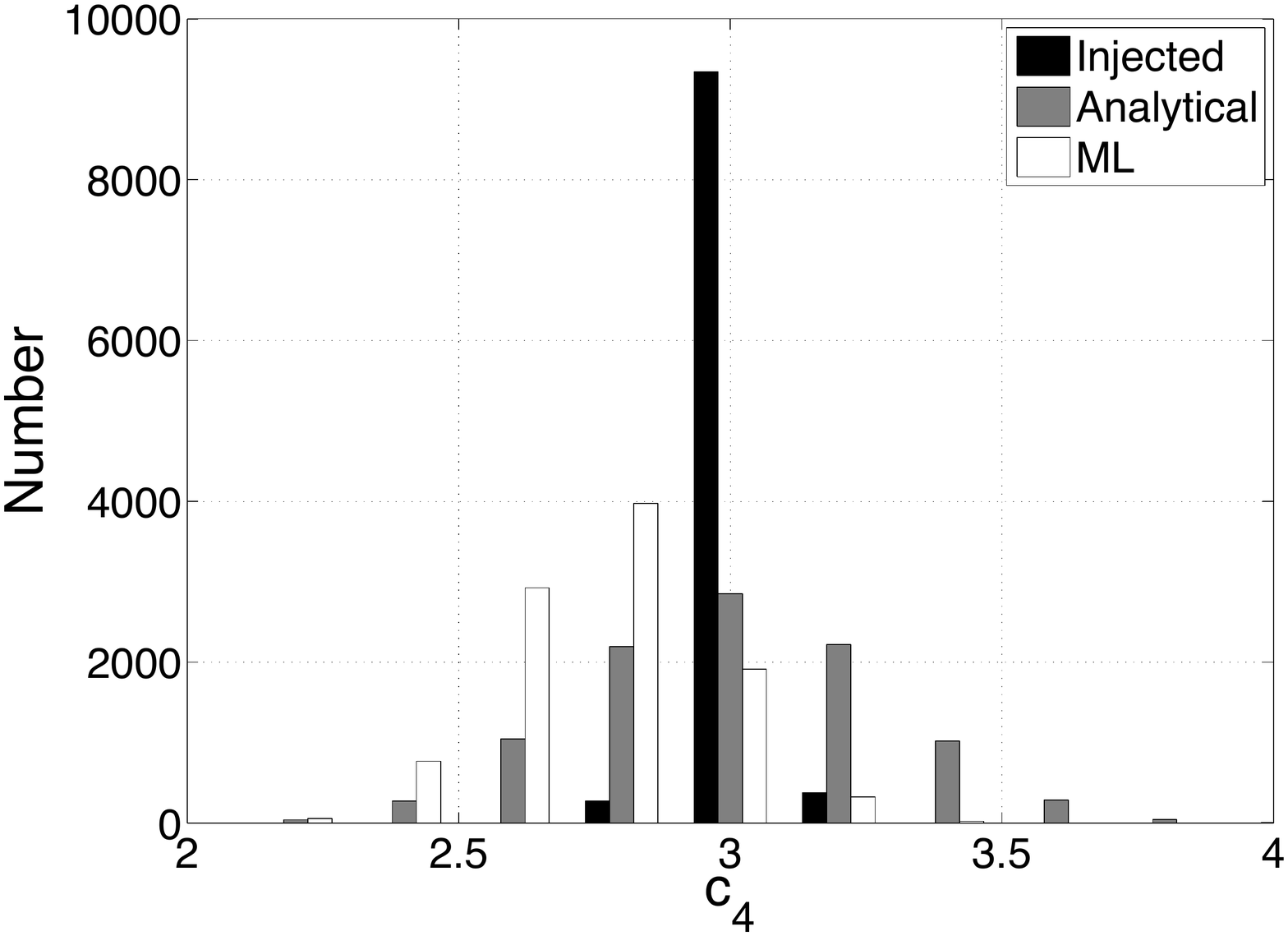}\\
 \caption{\label{fig2} Histogram of $c_4$ injected (black), estimated analytically (grey) and estimated numerically by maximizing the likelihood function (white),  for the distribution of Laplace (top left), Logistic (top right), hypersecant (bottom left) and NIG (bottom right). Each plot is the result of 10,000 realizations, each having $10^6$ points and with $\alpha^2 /\sigma_n^2=0.1$.  }
\end{figure}

In order to test our new likelihood statistic, we generate fictitious data sets $h_1(t)$ and $h_2(t)$ at the output of two co-incident detectors, containing the GW signal $s(t)$, with an outcome randomly selected from the distributions presented above, and independent Gaussian noises $n_1(t)$ and $n_2(t)$. 
We then use the simulated data to obtain analytically the moment-based estimates for $c_2$ and $c_4$ and also to obtain numerically the maximum likelihood estimates for these parameters. The results for $c_4$ averaged over $10^4$ trials for a number of point $T=10^6$ and for the 4 distributions are presented in Table ~\ref{tab1} and Figure ~\ref{fig2}.
The number of points in a sample containing 1 year of data sampled at about 100 Hz will be rather of the order of $10^9$ but this would have required more computational resources than what we had available. In order to get a realistic estimate of the performance, one should therefore divide by $\sqrt{10^3}$  the standard deviations quoted in the table. 
The number of trials has an effect on the average estimated value, especially when the standard deviation is large. In the limit, we expect that using 	n increasing number of trials would generate an estimate that would converge to the injected value in all cases. For this reason we only considered  values of the ratio $\alpha^2/\sigma_n^2$ larger than 0.03 (here we assume for simplicity that the variance of the noise is the same for both detectors, and we denote it by $\sigma_1=\sigma_2=\sigma_n$). With $\alpha^2/\sigma_n^2=0.01$, the uncertainty obtained with $10^6$ points is too large and we would need to average over more than $10^4$ trials to get a reasonably reliable estimate for$c_4$ . 
The results we obtain are very similar for the Logistic, Hypersecant and NIG distributions. The cumulant $c_2$ estimated with our new statistic, which we do not report here, is the same as the one derived analytically or obtained with the standard cross-correlation statistic, and is also in very good agreement with the injected value (better than 1\% for 1 year). This suggests that our methodology allows us to estimate $c_4$ accurately without any negative impact on how well estimated is the $c_2$ parameter. 
Note that the ratio $\alpha^2/\sigma_n^2 = 0.01-0.1$ is in the range of predicted values for cosmological and astrophysical stochastic backgrounds for both Advanced LIGO and VIRGO detectors and Einstein Telescope.
For cosmic strings, the typical value of the energy density parameter at 100 Hz is expected to be $\Omega_{gw}=10^{-9}-10^{-5}$ which corresponds to $\alpha^2/\sigma_n^2=10^{-6}-1$ for Advanced detectors and $\alpha^2/\sigma_n^2=10^{-4}-100$ for Einstein Telescope. For compact binary mergers, $\Omega_{gw}=10^{-10}-10^{-7}$ which corresponds to $\alpha^2/\sigma_n^2=10^{-7}-0.01$ for Advanced detectors and $\alpha^2/\sigma_n^2=10^{-5}-1$ for Einstein Telescope.

\begin{table}
\caption{\label{tab1} Average value over $10^4$ trials of $c_4$ estimated analytically and by likelihood maximization, for all the distributions considered in this paper, with standard deviation presented in parenthesis. The number of points in each trial is $10^6$. To obtain the estimation error for $10^9$ points, which  corresponds to a sample of 1 year of data, one should divide the standard deviation by $\sqrt(10^3)$.  
}
\begin{ruledtabular}
\begin{tabular}{llllll}
Distribution   && $\alpha^2 \sigma_n^{-2}=0.03$ & $\alpha^2 \sigma_n^{-2}=0.05$ & $\alpha^2 \sigma_n^{-2}=0.1$ \\

\hline
\textbf{Laplace} 
& anal &12.1 (9.6) & 12.0  (3.7) & 12.0 (1.1) \\
&ML & 12.1 (6.1) & 11.8 (2.4) & 11.3 (0.7)\\
\hline        
\textbf{Hypersecant }
& anal &2.0  (2.0) & 2.0 (0.9) & 2.0 (0.3)\\
&ML& 2.0 (1.4)  & 2.0 (0.6) & 2.0 (0.2)\\
\hline
\textbf{Logistic }
& anal & 1.2 (2.4) & 1.2 (0.9) & 1.2 (0.3) \\
& ML&1.4 (1.3) & 1.2 (0.6) & 1.2 (0.2)  \\
\hline
\textbf{NIG}
& anal & 3.0 (2.4) & 3.0 (0.9) & 3.0 (0.3) \\
& ML & 3.0 (1.5) & 3.0 (0.6) & 2.9 (0.2) \\
\end{tabular}
\end{ruledtabular}
\end{table}

We find that the fourth cumulant parameter is estimated with reasonably good precision for the cases under investigation. We also find that the maximum likelihood estimator generates an estimation error lower than that of the moment-based estimator, which is consistent with the general statement that the maximum likelihood estimator is asymptotically efficient.

\section{Extending the Approach to the Detection of Signals in the Popcorn
Regime}

A gravitational wave stochastic background is produced by a collection of
independent gravitational wave events. In some cases, the ratio of the
average time between events to the average duration of an event is small and
a large number of events are taking place simultaneously. In other cases,
the ratio is large and the signal received has a popcorn signature. In what
has been discussed so far in this paper, we have analyzed the first type of
situations, and implicitly considered that GW events were too numerous to be
individually distinguished, and yet not numerous enough for central limit
theorem to give a strictly Gaussian distribution, with a deviation explicitly characterized in terms of the Edgeworth expansion. We now turn to an analysis
of the second type of situations, following and generalizing an approach
introduced by DF03 \cite{2003PhRvD..67h2003D}, who have focused on a model where
the deviation from the Gaussian distributional assumption was understood as
emanating from the presence of a resolved Gaussian signal being measured
with a probability $0<\xi \leq 1$ (the Gaussian case is recovered for $\xi
=1 $). In DF03 \cite{2003PhRvD..67h2003D}, the observed distribution was
assumed to the of the following form: 
\begin{equation}
f_{s}\left( s_{t}\right) =\xi f_{G}\left( s_{t}\right) +\left( 1-\xi \right)
\delta \left( s_{t}\right) =\xi \frac{1}{\sqrt{2\pi }\alpha }e^{-\frac{%
s_{t}^{2}}{2\alpha ^{2}}}+\left( 1-\xi \right) \delta \left( s_{t}\right)
\end{equation}%
where the parameter $0<\xi \leq 1$ and $\delta \left( .\right) $ is the
density of the Dirac distribution. Since by assumption the events that are measured are supposed to be
in small numbers, the Gaussian assumption is hard to justify and should be
relaxed. We generalize this model by considering:%
\begin{equation}
f_{s}\left( s_{t}\right) =\xi f_{NG}\left( s_{t}\right) +\left( 1-\xi
\right) \delta \left( s_{t}\right)\end{equation}%
and we further assume that the unknown non-Gaussian density $f_{NG}$ can be
approximated by a 4th order Edgeworth expansion
\begin{eqnarray}
f_{s}\left( s_{t}\right) &=&\xi f_{G}\left( s_{t}\right) g\left(
s_{t}\right) +\left( 1-\xi \right) \delta \left( s_{t}\right) \\
&=&\xi \frac{1}{\sqrt{2\pi }\alpha }e^{-\frac{s_{t}^{2}}{2\alpha ^{2}}}\left[
1+\frac{c_{3}}{6\alpha ^{3}}H_{3}\left( \frac{x}{\alpha }\right) +\frac{c_{4}%
}{24\alpha ^{4}}H_{4}\left( \frac{x}{\alpha }\right) +\frac{c_{3}^{2}}{%
72\alpha ^{6}}H_{6}\left( \frac{x}{\alpha }\right) \right] +\left( 1-\xi
\right) \delta \left( s_{t}\right)
\end{eqnarray}

Using this expression for the signal density, we obtain the following
generalized form for the likelihood ratio:%
\begin{eqnarray*}
\Lambda _{ML} &=&\frac{\underset{\alpha ,c_{3},c_{4},\sigma _{1},\sigma
_{2},\xi }{\max }\int f_{s}\left( s\right) f_{n}\left( h-s\right) ds}{%
\underset{\sigma _{1},\sigma _{2}}{\max }f_{n}} \\
&=&\underset{\alpha ,c_{3},c_{4},\sigma _{1},\sigma _{2},\xi }{\max }%
\prod\limits_{t=1}^{T}\frac{\overline{\sigma }_{1}\overline{\sigma }_{2}}{%
\sigma _{1}\sigma _{2}}\int_{-\infty }^{+\infty }f_{s}\left( s_{t}\right)
\exp \left[ -\frac{\left( h_{1t}-s_{t}\right) ^{2}}{2\sigma _{1}^{2}}-\frac{%
\left( h_{2t}-s_{t}\right) ^{2}}{2\sigma _{2}^{2}}+1\right] ds_{t} \\
&=&\underset{\alpha ,c_{3},c_{4},\sigma _{1},\sigma _{2},\xi }{\max }%
\prod\limits_{t=1}^{T}\frac{\overline{\sigma }_{1}\overline{\sigma }_{2}}{%
\sigma _{1}\sigma _{2}}\int_{-\infty }^{+\infty }\left[ \xi f_{G}\left(
s_{t}\right) g\left( s_{t}\right) +\left( 1-\xi \right) \delta \left(
s_{t}\right) \right] \exp \left[ -\frac{\left( h_{1t}-s_{t}\right) ^{2}}{%
2\sigma _{1}^{2}}-\frac{\left( h_{2t}-s_{t}\right) ^{2}}{2\sigma _{2}^{2}}+1%
\right] ds_{t}
\end{eqnarray*}

After calculations similar to what has been done before for the case with $\xi=1$ , we obtain:%
\begin{eqnarray*}
\Lambda _{ML} &=&\underset{\alpha ,c_{3},c_{4},\sigma _{1},\sigma _{2},\xi }{%
\max }\prod\limits_{t=1}^{T}\left\{ \xi \frac{\sigma }{\alpha }\frac{%
\overline{\sigma }_{1}\overline{\sigma }_{2}}{\sigma _{1}\sigma _{2}}\exp %
\left[ -\frac{h_{1t}^{2}}{2\sigma _{1}^{2}}-\frac{h_{2t}^{2}}{2\sigma
_{2}^{2}}+1\right] \exp \left[ \frac{1}{2}\sigma ^{2}\left( \frac{h_{1t}}{%
\sigma _{1}^{2}}+\frac{h_{2t}}{\sigma _{2}^{2}}\right) ^{2}\right] \right. 
\notag \\
&&\left. \times \left( I_{0}+I_{1t}+I_{2t}+I_{3t}+I_{4t}+I_{6t}\right)
+\left( 1-\xi \right) \frac{\overline{\sigma }_{1}\overline{\sigma }_{2}}{%
\sigma _{1}\sigma _{2}}\exp \left[ -\frac{h_{1t}^{2}}{2\sigma _{1}^{2}}-%
\frac{h_{2t}^{2}}{2\sigma _{2}^{2}}+1\right] \right\}
\end{eqnarray*}

The values of $1-\xi ,$ $\alpha ,$ $c_{3},$ $c_{4},$ $\sigma _{1},$ $\sigma
_{2},$ that achieve the maximum value for the likelihood function are,
respectively, estimators for the probability of the presence of a (non-Gaussian) signal, the 2nd, 3rd
and 4th cumulant of the signal distirbutions, and the variance of the noise in the two
detectors. Note that if we evaluate this function at $\xi =1$, $%
c_{3}=c_{4}=0 $, rather than maximizing over $\xi ,$ $c_{3}$, $c_{4}$, we
recover the Gaussian detection statistic.

\section{Conclusions and Directions for Further Research}

This paper introduces a non-parametric approach to the detection of
non-Gaussian gravitational wave stochastic backgrounds. The approach we
propose is based on Edgeworth expansion, which is a formal expansion of the
characteristic function of the signal distribution, whose unknown
probability density function is to be approximated in terms of the
characteristic function of the Gaussian distribution. The non-Gaussian
detection statistic we obtain generalizes the standard cross-correlation
statistic that is recovered in the limit of vanishing third and fourth
cumulants of the empirical conditional distribution of the detector
measurement. Our paper complements related work by DF03 \cite {2003PhRvD..67h2003D}, who have focused on a very specific model where the deviation from
the Gaussian distributional assumption was understood as emanating from the
presence of a resolved Gaussian signal being measured. We provide a methodology
that can be applied without any assumption regarding the exact nature of the
departure from normality, and which relies on an explicit correction to the central limit theorem, when the number of sources is finite. 
The main benefit of the procedure is that it allows us to
estimate additional parameters when the signal is not too small compared to the noise ($\alpha^2/\sigma_n^2$ of the order of 0.01), namely the 3rd and 4th cumulant of the
gravitational wave signal distribution, which should be useful for analyzing
the constraints on astrophysical and cosmological models that will be
imposed by observed gravitation wave signals, and comparing them to the
constraints derived from supernovae or galaxy clusters observations. Our
methodology can be extended to a situation when there is uncertainty about
the signal presence, e.g., in the case of bursts, where there is a positive
probability of having no signal, and having few of them simultaneously when
they are there, which would allow us to generalize the methodology in DF03 \cite{2003PhRvD..67h2003D} to a non-Gaussian signal.

More generally, we may also account for the presence of a random,
exponentially-distributed, number of sources, each of which having a
non-Gaussian distribution, by using formal expansions for compound Poisson
processes (see for example \cite{babu2003edgeworth}). Our approach is also
flexible enough to be extended to the presence of non-Gaussian noise
distributions, in addition to the presence of non-Gaussian signal
distribution. This extension would be useful since significant non-Gaussian
tails have been obtained for the noise distributions of all gravitational
waves detectors (see \cite{2003PhRvD..67l2002A}) and will be the purpose of a future paper.

Finally, in this paper we considered coincident detectors. This assumption is valid for Einstein Telescope but more work is needed to extend our results to a network of separated detectors such as  LIGO/Virgo network. 

\bibliography{biblio}
\end{document}